\documentclass[amsmath,amssymb,aps,twocolumn,pra]{revtex4-2}
\usepackage{graphicx}% Include figure files%
\usepackage{dcolumn}% Align table columns on decimal point
\usepackage{bm}% bold math
\usepackage{hyperref}% add hypertext capabilities
\usepackage{multirow} % for cmd 'multirow', 'multicolumn'
\hypersetup{colorlinks=true, citecolor=blue, urlcolor=blue, linkcolor=blue}

\begin{document}

\preprint{APS/123-QED}
\date{\today }
\title{Quantum tricriticality and universal scaling in a tricritical quantum Rabi system}
\author{You-Qi Lu}
\affiliation{Department of Physics, and Chongqing Key Laboratory for strongly coupled Physics,  Chongqing University, Chongqing 401330, China}
\author{Yu-Yu Zhang}
\email{yuyuzh@cqu.edu.cn}
\affiliation{Department of Physics, and Chongqing Key Laboratory for strongly coupled Physics,  Chongqing University, Chongqing 401330, China}

\begin{abstract}
Quantum tricriticality, a unique form of high-order criticality, is expected to exhibit fascinating features including unconventional critical exponents and universal scaling laws. However, a quantum tricritical point (QTCP) is much harder to access, and the corresponding phenomena at tricriticality have rarely been investigated. In this study, we explore a tricritical quantum Rabi model, which incorporates a nontrivial parameter for adjusting the coupling ratio between a cavity and a three-level atom. The QTCP emerges at the intersection of a first- and second-order superradiant phase transitions according to Landau theory. By using finite-frequency scaling analyses for quantum fluctuations and the mean photon number, universal critical exponents differentiate the QTCP from the second-order critical point. We find that the phase transition at the tricritical point goes beyond the conventional second-order phase transition. Our work explores an interesting direction in the generalization of the well-known Rabi model for the study of higher-order critical points due to its high control and tunability.
\end{abstract}

\maketitle

\textit{Introduction --}Quantum phase transition (QPT) is a central issue in the study of many-body quantum phenomena at zero temperature~\cite{Sachdev}. Characterizing universal phase transition phenomena and identifying critical exponents are essential in understanding phase transitions. Quantum critical points are often observed as a divergence point of an order parameter in continuous phase transitions by adjusting external physical parameters such as magnetic fields~\cite{RevModPhys2007,RevModPhys.69.315}. In contrast to conventional critical points, a quantum tricritical point (QTCP) arises where a continuous phase transition changes into a discontinuous one.  QTCPs were originally
found in He$^{3}$-He$^{4}$ mixtures in finite
temperature phase diagrams, which were characterized by the Landau theory of phase transitions~\cite{griff}. Tricriticality is challenging to access
in real materials, but it can be found, for example, in itinerant ferromagnets~\cite{PhysRevLett1999} and metallic magnets~\cite
{Belitz,Canfield,yuan2019,Friedemann}. Several experimental
and theoretical works indicate unconventional quantum criticalities resulting from quantum tricriticalities in many-body systems~\cite{Friedemann,pu,PhysRevLett2018,PhysRevB2015,zhang2022,PhysRevA.108.033706} .

QPTs in light-matter interaction systems have been extensively studied in recent years, leading to the discovery of exotic quantum phases in quantum many-body systems~\cite{Greentree2006,plenio,zhu2020}.  One well-known quantum phenomenon is the superradiant phase transition, which occurs when a collection of two-level atoms undergoes spontaneous emission~\cite{Dicke,lambert,chen2008,zhang2019,Emary03}, This phenomenon has been observed in Bose–Einstein condensate gas~\cite{nagy} and degenerate Fermi gas experiments~\cite{sciencewu}. The Quantum Rabi model, consisting of a two-level system and a bosonic field mode, also exhibits a superradiant phase transition in a infinite-frequency ratio limit with analogy to a thermodynamic limit~\cite{Ashhab2013,PhysRevLett.115.180404,PhysRevLett.119.220601,PhysRevA.101.033827,PhysRevApplied.9.064006}. This has been achieved in quantum simulations~\cite{chen2021,NCcai2021}. Significant efforts have been dedicated to exploring the existence of QPTs in few-body systems in finite Jaynes-Cummings lattice systems~\cite{PhysRevLett.117.123602},  anisotropic quantum Rabi and Rabi-star model~\cite{PhysRevA.104.043307,PhysRevA.102.063721,PhysRevA.95.013819}, and quantum Rabi ring with an artifical field~\cite{PhysRevLett.127.063602,PhysRevLett.129.183602,zhang2023}. The QPTs in a few-body system offer an avenue for investigating nontrivial criticality and exotic phases due to high degree of tunability.

In this study, we explore a generalization of the well-known Rabi model, aiming to find profound high-order criticality. We introduce a tricritical quantum Rabi model that incorporates a tunable parameter that gauges the ratio between the coupling strengths of the cavity and a three-level atom. By using Landau theory, we expand the ground-state energy in terms of an order parameter, revealing superradiant phase transitions of first and second orders. Notably, a QTCP emerges at the boundary between critical lines for first- and second-order quantum phase transitions. At the QTCP, the scaling exponents of quantum fluctuations and the mean photon number are different from exponents at the second-order critical point. Our findings indicate that the QTCP belongs to a distinct universality class with a unique universal critical exponent, differing from that of the conventional quantum Rabi model.

\begin{figure}[htbp]
	\centering
	\includegraphics[width=0.45\textwidth]{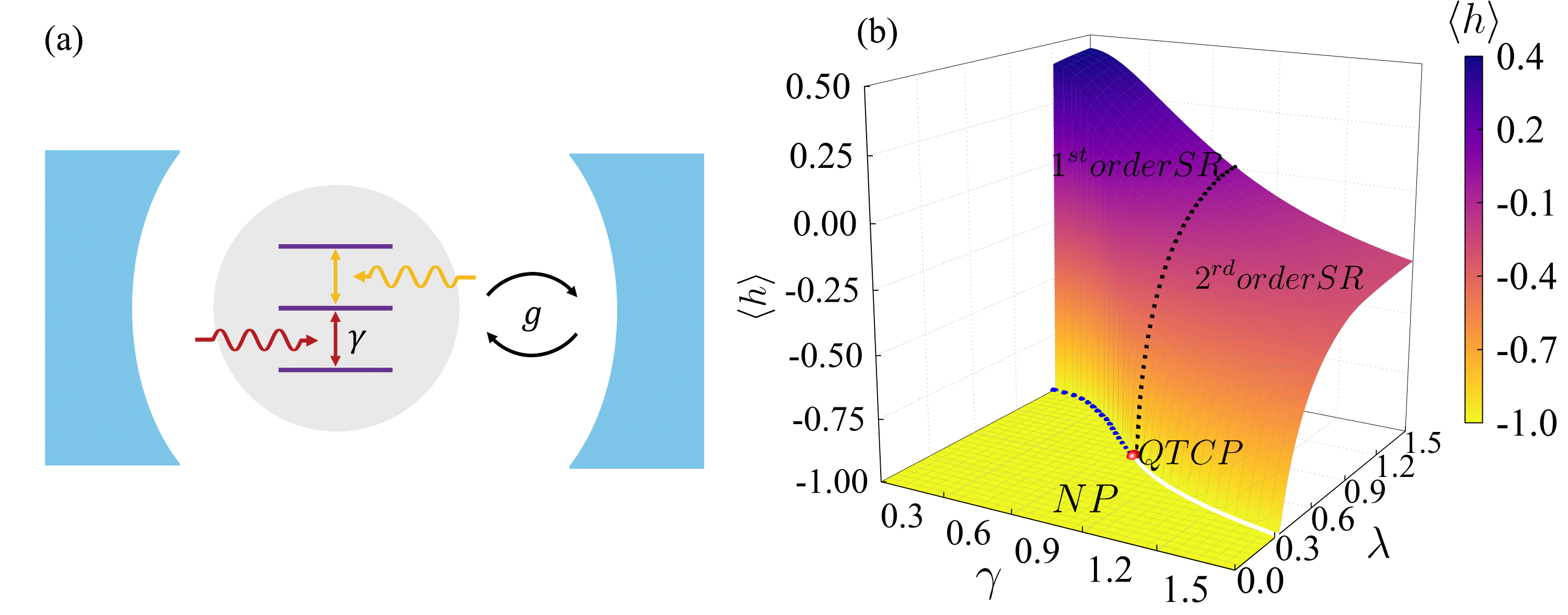}
	\caption{(a) An setup of the tricritical Rabi system. A three-level atom is coupled to a single-mode cavity. Three levels of the atom are coupled by cavity-assisted Raman transitions, for which the atomic transition ratio $\gamma$ is tuned from the side by two driving lasers in red and yellow lines. (b) Average value of the atom energy $\langle h\rangle$ in the $\gamma - \lambda$ plane for the phase transitions from the NP ($\lambda<\lambda_c$) to the second-order SR and first-order SR phases ($\lambda>\lambda_c$), respectively. The white solid line is a second-order critical line while the blue dashed line is a first-order critical line, respectively. The QTCP is marked by a red dot. In all our calculations, we set $\omega=1$ as the units for
frequency.} 
	\label{Emf}
\end{figure}
\textit{Tricritical quantum Rabi model --}We consider a tricritical quantum Rabi system, which describes a three-level atom coupled uniformly to a single-mode cavity. The Hamiltonian of this system is a generalization of the well-known quantum Rabi model and reads
\begin{equation}\label{Hamiltonian}
	H_{R} = \omega a^{\dagger}a + g( a^{\dagger} + a )d + \Omega h,
\end{equation}
where $a(a^{\dagger})$ denotes the photon annihilation~(creation) operator of the single-model cavity with the frequency $\omega$. $g$ is the atom-cavity coupling strength, $\Omega$ characterizes the atom energy splitting. The dipole operator $d$ of the atom and the single-atom Hamiltonian $h$ are defined as
\begin{equation}\label{d&h}
d=\left(\begin{array}{ccc}
	0 & 1 & 0 \\
	1 & 0 & \gamma \\
	0 & \gamma & 0
\end{array}\right), ~~~~~ h=\left(\begin{array}{ccc}
	1 & 0 & 0 \\
	0 & 0 & 0 \\
	0 & 0 & -1
\end{array}\right).
\end{equation}
The dipole operator $d$ incorporates a nontrivial parameter $\gamma$ that tunes the strength ratio of the atomic transitions between $\left |1\right \rangle \leftrightarrow\left |0\right \rangle $ and $\left |0\right \rangle\leftrightarrow \left |-1 \right \rangle $, for which $\left |\varepsilon_i\right \rangle$ ($\varepsilon_i=0,\pm1$) is the eigenstates of $h$ of the three-level atom. $\gamma$ plays a crucial influence on the effective coupling strength between the cavity and the atom. For a experimental realization in Fig.~\ref{Emf} (a), a three hyperfine levels could be the ones on the F$= 1$ ground state of $87$Rb. Three levels are coupled through the cavity and laser fields, which can be realized by cavity-assisted Raman transitions~\cite{PhysRevLett.119.213601}. The three-level atom interacts with two coupling laser with different frequencies, which control the atomic transitions ratio $\gamma$.  In the following we denote a scaled dimensionless coupling strength as $\lambda = g / \sqrt{\Omega \omega}$.  

For a weak atom-cavity coupling $\lambda$, the excitation tends to zero, which corresponds to the normal phase (NP). As $\lambda$ increases to a critical value $\lambda_c$, the photon population becomes macroscopic, and the system enters a superradiant (SR) phase. The SR phase transitions occurs in the infinite-frequency limit by denoting $\eta =\Omega/\omega\rightarrow \infty$, which is analogous to the infinite limit in the quantum Rabi model~\cite{PhysRevLett.115.180404,PhysRevLett.119.220601}.

\textit{Superradiant phases and a tricritical point --}In the superrdiant phases, the excitation is proportional to $\eta$ due to macroscopic population. Then we follow a mean-field appraoch by shifting the bosonic operator with respected to their mean value, $a\rightarrow a+\beta$ with $\beta=\langle a\rangle\propto\sqrt{\eta}$. The effective ground-state energy term of the Hamiltonian in Eq.(\ref{Hamiltonian}) is obtained as 
\begin{eqnarray}\label{MF}
H_0/\Omega&=& h+\frac{2\lambda}{\sqrt{\eta}}\beta d+\frac{1}{\eta} \beta^{2},
\end{eqnarray}
where $h$ and $d$ are the single atom operators given in Eq.~(\ref{d&h}). The ground-state energy is determined by a nonzero value of $\beta$ minimizing the energy of $H_0$ term. Fig.~\ref{Emf} (b) displays the average value $\langle h\rangle$ dependent on $\gamma$ and $\lambda$, which is obtained by numerical variational method. In the NP regime with the coupling strength $\lambda<\lambda_c$, $\langle h\rangle$ equals to $-1$ corresponding to the atom in the lowest state.  As $\lambda$ exceeds the critical value $\lambda_c$, $\langle h\rangle$ smoothly increases from $-1$, revealing a second-order SR phase transition. On the other hand, when  $\gamma$ is below a critical value $\gamma_{\texttt{TCP}}$,  $\langle h\rangle$ changes sharply with a discontinuous increasing, manifesting a first-order SR phase. Moreover, the critical lines of the first and second-order phase transitions intersect at a QTCP ($\gamma_{\texttt{TCP}},\lambda_{\texttt{TCP}}$) in a red dot. We analyze the emergence of the first and second-order phase transitions in the following using the Landau theory approach.

\begin{figure}
	\centering
	\includegraphics[width=0.45\textwidth]{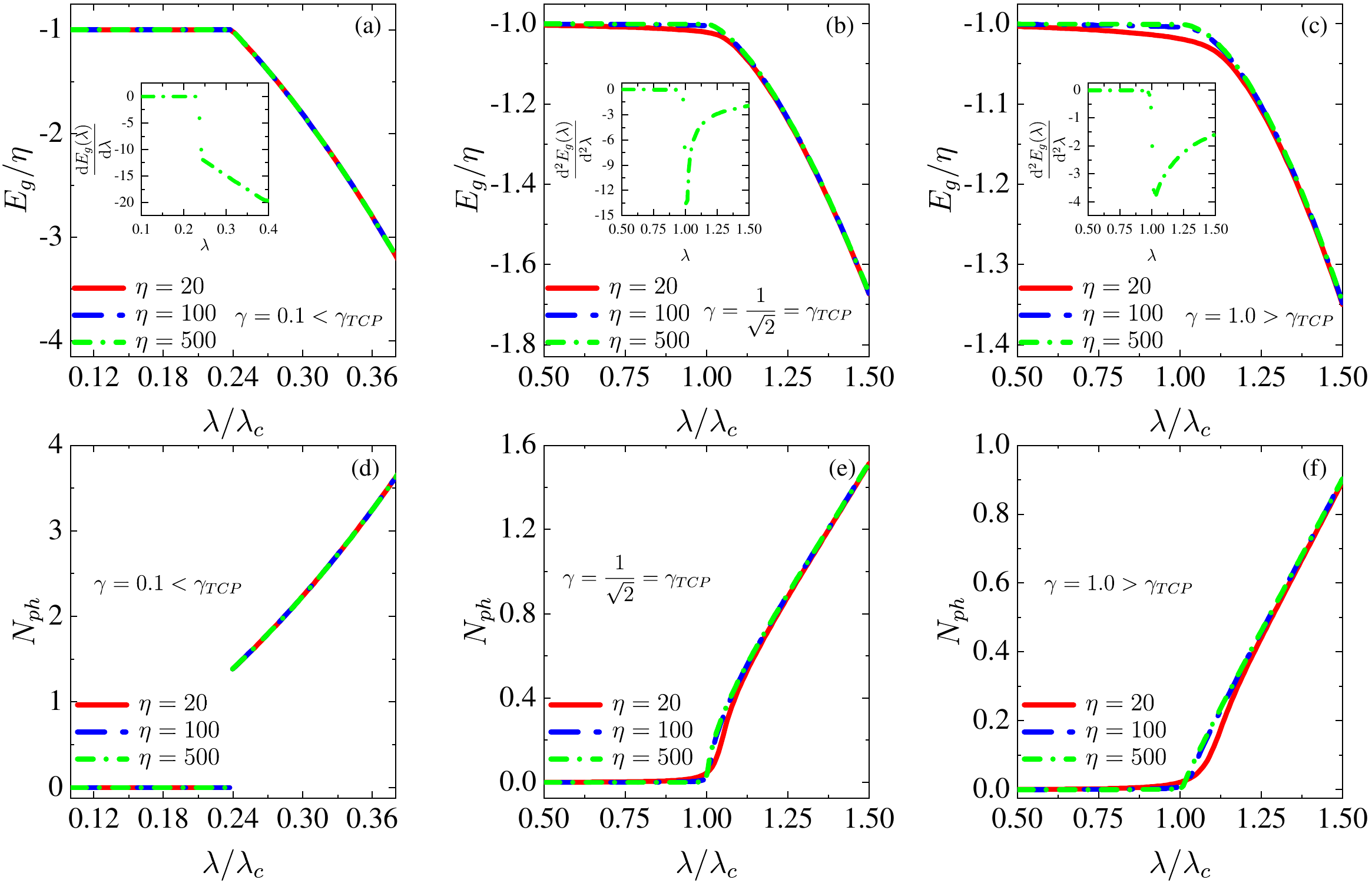}
	\caption{(a-c) Scaled ground-state energy $E_g/\eta$ and the mean photon number $N_{ph}$ as a function of the dimensionless coupling strength $\lambda/\lambda_{c}$ for $\gamma = 0.1$ (a) and (d),  $\gamma =\gamma_{\texttt{TCP}}= 1/\sqrt{2}$ (b) and (e), and  $\gamma = 1.0$ (c) and (f) for $\eta=20,100,500$. The first and second derivatives of the energy with respected to $\lambda$, $dE_{g}/d\lambda$ and $d^2E_{SR}/d^2\lambda$, are listed in the insets. }
	\label{Occ-Eg}
\end{figure}

Since the ground-state energy term in Eq.(\ref{MF}) can be reduced to $H_0/\Omega=h+\alpha d+\alpha^{2}/4\lambda^{2}$ with the rescaled order parameter $\alpha= 2\lambda\beta/\sqrt{\eta}$.
In the infinite-frequency limit $\eta \rightarrow \infty$, the ground-state energy of the effective Hamiltonian (\ref{MF}) can be expanded as a Taylor series in terms of the tiny value of $\alpha$, $E_{SR} = \sum_{k=0}^{\infty} c_{k} \alpha^{2k}$. The coefficients $c_{k}$ are obtained using the perturbation theory by treating $h$ as the unperturbed Hamiltonian and $\alpha d$ term in Eq. (\ref{MF}) as the perturbation. Considering the Landau theory, we perform the sixth-order perturbation energy by keeping the expansion up to order $\alpha^{6}$ as
\begin{equation}\label{mfenergy}
\frac{E_{SR}}{\Omega} = c_{1} \alpha^{2} + c_{2}\alpha^{4} + c_{3} \alpha^{6}-1,
\end{equation}
where the coefficients $c_{1} = 1/(4 \lambda^{2}) - \gamma^{2}$,~$c_{2} = \gamma^{2}(\gamma^{2} - \frac{1}{2})$, and~$c_{3} = -\gamma^{2}(1-7\gamma^{2} + 8\gamma^{4})/4$ are given in the Appendix. According to the first derivatives of $E_{SR}$ with respective to $\alpha$, $dE_{SR}/d\alpha=0$, we obtain minimum values
\begin{eqnarray}\label{solution}
\alpha_{\pm} = \pm \sqrt{(-c_{2} + \sqrt{c_{2}^{2}-3c_{1}c_{3}}) / 3 c_{3}},
\end{eqnarray}
and $\alpha=0$. Obviuosly, the ordinary second-order critical boundary is obtained when $c_{1} = 0$ and $c_{2} > 0$. It results in the second-order critical boundary
\begin{eqnarray}\label{critical value}
\lambda_{c}= \frac{1}{2\gamma}.
\end{eqnarray}
It fits well with the critical line of the second-order phase transition in Fig.~\ref{Emf}(b) in white solid line.

The QTCP, where marks the intersection of first- and second-order phase transitions, is determined from  $c_{1} = c_{2} = 0$ and $c_{3} > 0$. It yields the tricritical point 
\begin{equation}
\gamma_{\texttt{TCP}} = 1 / \sqrt{2},\lambda_{\texttt{TCP}} = 1 / \sqrt{2}.
\end{equation} 
The location of the QTCP is marked in a red dot in Fig.~\ref{Emf}(b).

When $\gamma \ge \gamma_{\texttt{TCP}}$, the ground-state energy $E_{SR}$ has two global minima at $\alpha_{\pm}$ with the coefficient $c_{1}<0$, signaling the second-order phase transition. For $\gamma < \gamma_{\texttt{TCP}}$, $E_{SR}$ has three local minima at $\alpha_{\pm}$ and  $\alpha = 0$. As the phase transition is crossed, the global minimum switches from $\alpha = 0$ to $\alpha_{\pm}$. The discontinuous jump of the energy in the global minimum location indicates a first-order phase transition in Appendix.

\begin{figure}
	\centering
	\includegraphics[width=0.45\textwidth]{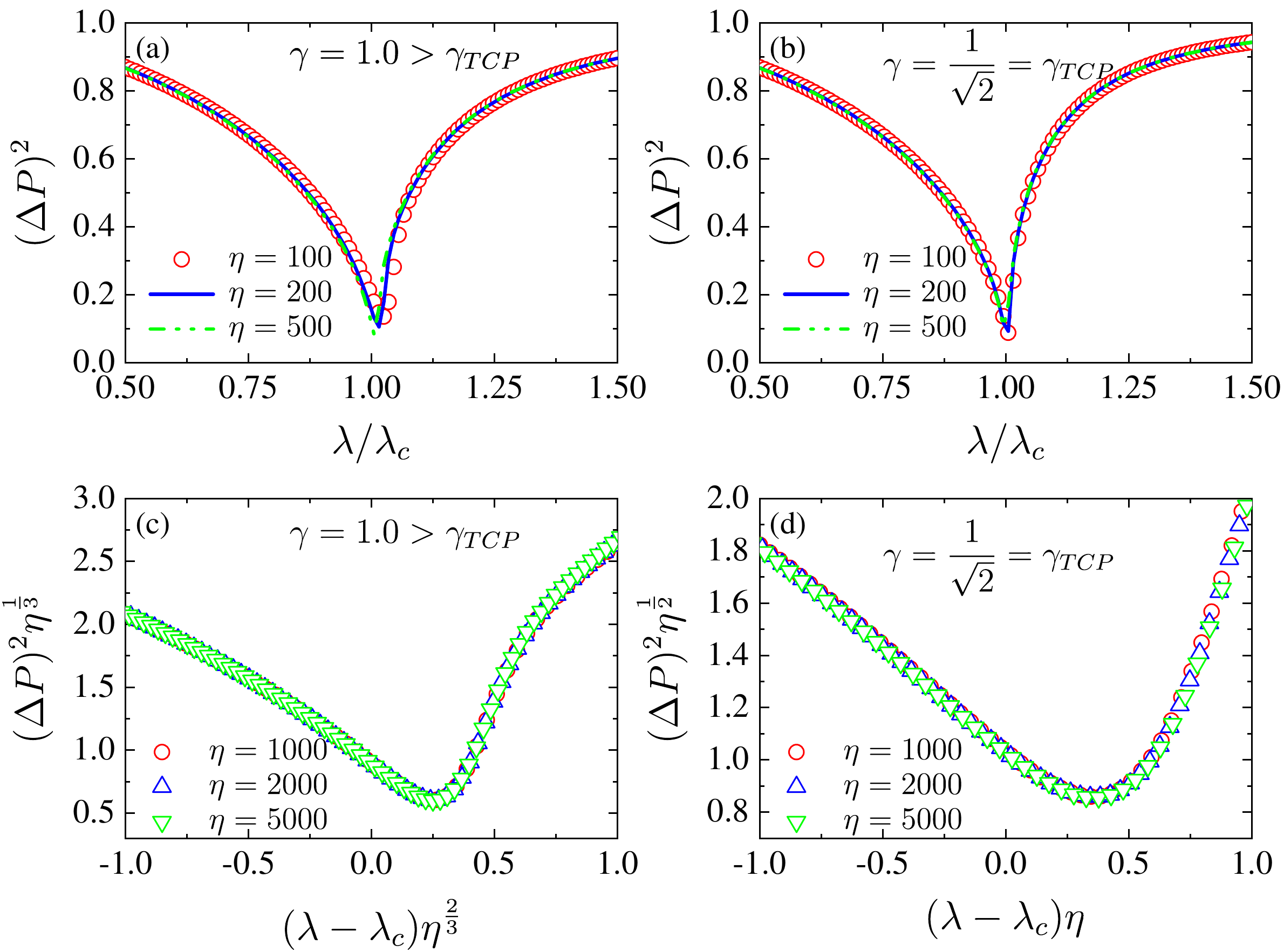}
	\caption{The variance of momentum $(\Delta P)^2$ as a function of the dimensionless coupling strength $\lambda/\lambda_{c}$ for $\gamma=1$ (a) and $\gamma=\gamma_{\texttt{TCP}}$(b) for finite frequency ratio $\eta=100,200,500$. Finite-frequency scaling functions for $\gamma=1$ (c) and $\gamma=\gamma_{\texttt{TCP}}$ (d) for large values $\eta=1000,2000,5000$.}
	\label{ScalingP}
\end{figure}

To show the validity of the perturbation theory, we accurately calculate the scaled ground-state energy $E_g/\eta$ and the scaled mean photon number $N_{ph}=\langle a^{\dagger}a\rangle/\eta$ of the Hamiltonian (\ref{Hamiltonian}) by numerical exact diagonalization. In the NP ($\lambda/\lambda_c<1$), the excitation of photons $N_{ph}$ tends to zero due to zero excitation, while it increases in the superradiant phase ($\lambda/\lambda_c>1$). For the transition strength ratio $\gamma=0.1<\gamma_{\texttt{TCP}}$ in Fig.\ref{Occ-Eg} (a) (d), both $dE_g(\lambda)/d\lambda$ and $N_{ph}$ are discontinuous at the critical coupling strength $\lambda=\lambda_c$, revealing the first-order nature of the QPT. When $\gamma=1>\gamma_{\texttt{TCP}}$ in Fig.\ref{Occ-Eg} (c) (f), $N_{ph}$ becomes continuous, while $d^2E_g(\lambda)/d^2\lambda$ is discontinuous, indicating a second-order phase transition. The first- and second-order SR phase transitions are consistent with the analysis using the Landau theory. At the QCTP, $\gamma_{\texttt{TCP}}=1/\sqrt{2}$, Fig.\ref{Occ-Eg} (b) (e) show smooth behavior of both $E_g$ and $N_{ph}$. Moreover, the critical point approaches to the critical value $\lambda_c$ in Eq. (\ref{critical value}) when $\eta$ grows from $20$ to $500$, demonstrating the finite-frequency effect. 

\begin{figure}
	\centering
	\includegraphics[width=0.45\textwidth]{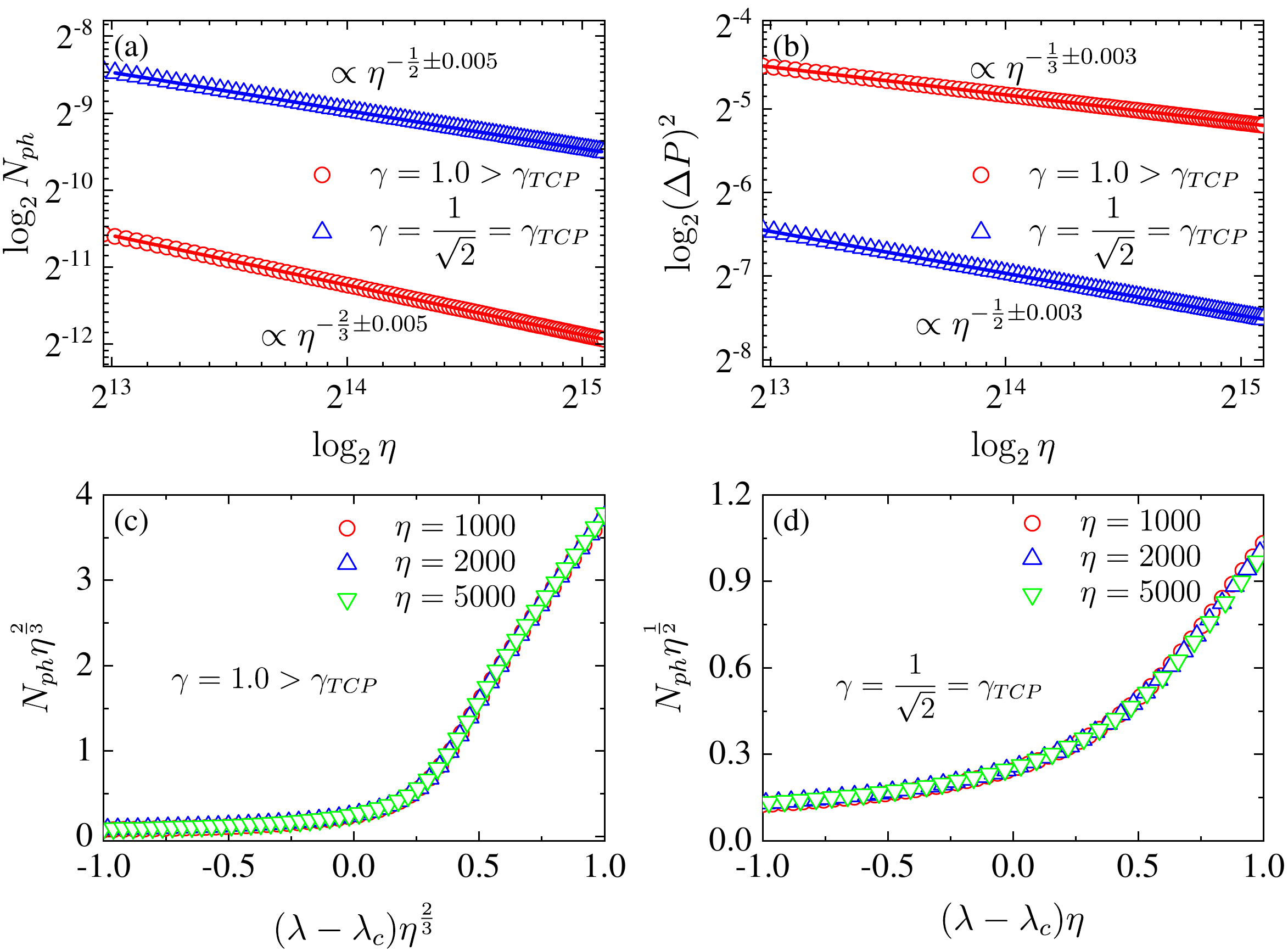}
	\caption{Scaling of the average photon number $N_{ph}$ (a) and variance of momentum $(\Delta P)^2$ (b) as a function of $\eta$ in log-log scale at the $2$nd-order critical point  with $\gamma = 1.0 > \gamma_{c}$ (red circles) and the QTCP with $\gamma=\gamma_{\texttt{TCP}}$ (blue triangle).  Finite-size scaling of the average photon number around the $2$nd-order critical point (c) and around the QTCP (d), respectively. }
	\label{ScalingNph}
\end{figure}

\renewcommand{\tabcolsep}{0.4cm}
\renewcommand{\arraystretch}{1.3}
\begin{table*}[]\centering 
	\centering
 \caption{Various critical exponents $\beta_Q$, $\beta_Q/\nu$ and $\nu$ obtained using the finite-frequency scaling function for the variance of momentum $\Delta P$, and the average photon number $N_{ph}$ for the QTCP and $2$nd-order critical point, respectively.}\label{table}
	\begin{tabular}{|cc|c|c|c|}
		\hline
		\multicolumn{2}{|l|}{}                                                                 & \begin{tabular}[c]{@{}c@{}}Critical exponent\\ $\beta_Q$\end{tabular} & \begin{tabular}[c]{@{}c@{}}Finite-frequency scaling \\ exponent $\beta_Q/\nu$\end{tabular} & Universal critical exponent $\nu $ \\ \hline\hline
		\multicolumn{1}{|c|}{\multirow{2}{*}{$(\Delta P)^{2}$}} & $2^{nd}$ order critical point
		&1/2               &1/3               & 3/2                   \\ \cline{2-5} 
		\multicolumn{1}{|c|}{}                            & QTCP 
		& 1/2              & 1/2             & 1                     \\ \hline
		\multicolumn{1}{|c|}{\multirow{2}{*}{$N_{ph}$}}   & $2^{nd}$ order critical point       
		& 1                & 2/3              & 3/2                   \\ \cline{2-5} 
		\multicolumn{1}{|c|}{}                            & QTCP
		& 1/2              & 1/2              & 1                     \\ \hline
	\end{tabular}
\end{table*}

\textit{Universal scaling and critical exponents --}To gain universal features of different phase transitions, it is of great interest to explore the critical exponents and universality classes of the QTCP and the second-order QCP in the tricritical Rabi model.
It is well known that different systems can exhibit similar quantum criticality, giving rise to universality. Finite-size scaling is a topic of major interest in QPT systems and has been firmly established since the development of a general theory~\cite{PhysRevLett.28.1516,PhysRevLett.49.478,Sachdev}. 

The finite-size scaling ansatz for a physical quantity $Q$ in the critical
region takes the following scaling law form 
\begin{equation}\label{scalingfunction}
Q(\eta, \lambda) = \eta ^{-\beta_{Q}/\nu} F_{Q}(|\lambda - \lambda_{c}|\eta^{1/\nu}),
\end{equation}
where $\nu$ is a universal critical exponent but independent of the physical quantity, $F_{Q}(x)$ is the scaling function of $Q$, and $\beta_{Q}$ is the critical exponent for $Q$.  The scaling form dependent on $\eta$ is similar to finite-size scaling in the thermodynamics phase transitions, which is known as finite-frequency scaling. 

At the critical point $\lambda_c$, one obtains log-log relation as 
\begin{equation}
\texttt{ln} Q(\eta, \lambda_c) =-\frac{\beta_{Q}}{\nu} \texttt{ln}\eta+\texttt{ln} F_{Q}(0),
\end{equation}
where $\texttt{ln} F_{Q}(0)$ is a constant. The critical exponent $\beta_{Q}/\nu$ is obtained as the slope of the linear dependence. It yields the finite-frequency scaling relation as $Q(\eta,\lambda_{c}) \propto \eta ^{-\beta_{Q}/\nu}$.

We consider the observable $Q$ as the variance $(\Delta P)^2=\langle P^2\rangle-\langle P\rangle^2$ of the momentum quadrature $P = i(a^{\dagger} -a)$, which is account for the quantum fluctuations. Fig.~\ref{ScalingP} (a)(b) shows the dependence of the $\Delta P$ on the coupling strength $\lambda$. As $\eta$ increases, the quantum fluctuations become divergent around the critical point. The finite-frequency scaling function for $\Delta P$ is calculated dependent on $\eta$ in a log-log relation, as illustrated in Fig.~\ref{ScalingNph} (b). From the slope of the lines, the critical exponent for the tricritical point is equal to $\beta_Q/\nu=1/2$, but it equals to $1/3$ for the $2$nd-order critical point. Thus, various finite-frequency scaling laws are obtained as  $(\Delta P)^2(\eta,\lambda_{c}) \propto \eta ^{-1/3}$ for the $2$nd-order critical point and $(\Delta P)^2(\eta,\lambda_{\texttt{TCP}}) \propto \eta ^{-1/2}$ for the QTCP, respectively. 

The scaling function in Eq.(\ref{scalingfunction}) should be universal for large $\eta$ at the critical regime, which is independent of $\eta$. According to the critical exponent $\beta_Q/\nu=1/3$ for the second-order phase transition, Fig.~\ref{ScalingP} (c) shows the universal scalings of $(\Delta P)^2\eta^{1/3}$ as a function of $(\lambda-\lambda_c)\eta^{1/\nu}$ for different $\eta$. Remarkably, an excellent collapse in the critical regime is observed according to the scaling function in the curve for $\eta=1000,2000, 5000$.  It demonstrates that the universal critical exponent is $\nu=3/2$ for the second-order phase transition,  which is the same as that in the quantum Rabi model and Dicke model~\cite{PhysRevLett.119.220601,PhysRevA.80.023810}. Meanwhile for the QTCP, the universal scaling function $(\Delta P)^2\eta^{1/2}$ as a function of $(\lambda-\lambda_{\texttt{TCP}})\eta^{1/\nu}$ is shown in Fig.~\ref{ScalingP} (d). One observs that curves with the universal critical exponent $\nu=1$ collapse together. Thus the universal scaling function of $\Delta P$ at the QTCP and the $2$nd-order critical point are obtained explicitly as 
\begin{align}
	& (\Delta P)^2(\eta, \lambda \to \lambda_{c})  \propto  \eta^{-1/3}F_{\Delta P}(|\lambda - \lambda_{c}|\eta^{2/3}), \\[4mm]
	& (\Delta P)^2(\eta, \lambda \to \lambda_{\texttt{TCP}})  \propto  \eta^{-1/2}F_{\Delta P}(|\lambda - \lambda_{\texttt{TCP}}|\eta). 
\end{align}
It demontrates that the universal exponent at the QTCP $\nu=1$ is different from $\nu=3/2$ at the second-order critical point.

To show the universal critical exponent $\nu$ independent of observables, we investigate the universal scaling of the average photon number $N_{ph}$. Figs.\ref{ScalingNph} (a) shows $N_{ph}$ as a function of $\eta$ in a log-log scale. 
The slope of the line at the QTCP gives the critical exponent $\beta_{Q}/\nu=1/2$, which is different from $\beta_{Q}/\nu=2/3$ at the $2$nd-order critical point. Around $2$nd-order critical point, curves of the scaling function for different scales of
$\eta$ collapse into a single curve  in Fig. \ref{ScalingNph}(c), which gives the universal critical exponent $\nu=3/2$. Around the QTCP, we calculate the universal scaling function $N_{ph}\eta^{1/2}$ dependent on $(\lambda-\lambda_{\texttt{TCP}})\eta^{1/\nu}$ in Fig. \ref{ScalingNph}(d). An collapse with $\nu=1$ is achieved for different $\eta$. Thus, for the average photon number $N_{ph}$, the scaling functions around the $2$nd-order critical point and the QTCP are obtained, respectively
\begin{align}
	& N_{ph}(\eta, \lambda \to \lambda_{c})  \propto  \eta^{-2/3}F_{N_{ph}}(|\lambda - \lambda_{c}|\eta^{2/3}), \\[4mm]
	&  N_{ph}(\eta, \lambda \to \lambda_{\texttt{TCP}})  \propto  \eta^{-1/2}F_{N_{ph}}(|\lambda - \lambda_{\texttt{TCP}}|\eta). 
\end{align}

Based on the universal scaling analysis, we have successfully captured various critical exponents that govern two types of phase transitions. Tab.~\ref{table} presents the critical exponents obtained using the finite-frequency scaling function. The critical exponent $\beta_Q$ varies for different observables $N_{ph}$ and $\Delta P$. In contrast, the critical exponent $\nu$ is a universal constant that is independent of the physical quantity. Both $\Delta P$ and $N_{ph}$ predicts the same value of $\nu=3/2$ for the $2$nd-order critical point. In comparison, the universal critical exponent at the QTCP is equal to $\nu=1$. It indicates that the QTCP belongs to a nontrivial universality class with different critical exponents, which goes beyond the second-order superradiant phase transition in the conventional quantum Rabi and Dicke models. Recently, a chiral tricritical point has also exhibited a distinct universality class of phase transitions~\cite{PhysRevLett2018,PhysRevLett.129.183602}. It demonstrate that it is nontrivial to explore quantum critical phenomenons at the tricritical points.

\textit{Conclusions --}In summary, we have investigated the first and second-order superradiant phase transitions in the tricritical quantum Rabi model. According to Landau theory, the ground-state energy is obtained up to sixth-order perturbation, which displays local minima for the first and second-order phase transitions. The tricritical point arises at the intersection of the boundaries for the first and second-order phase transitions. We perform finite-frequency scaling analysis to calculate the universal scaling of observables. We find the superradiant phase transition at the tricritical point belongs to a different universality class with a different universal critical exponent. The generalization of the well-known quantum Rabi model can serve as an valuable platform for exploring critical phenomena and more intricate critical behaviors in few-body systems.

\textit{Acknowledgments --}This work was supported by NSFC under Grant No.12075040 and No. 12347101, and Chongqing NSF under Grants No.cstc2020jcyj-msxmX0890.

\appendix
\section{Ground-state energy derived by perturbation theory}\label{appendix1}

We employ perturbation theory to derive the ground-state energy in Eq.~(\ref{mfenergy}) and the corresponding coefficients $c_k$ in the superradiant phase. The ground-state energy term of Hamiltonian $H_0$ (\ref{MF}) is rewritten as 
\begin{equation}\label{effham}
\frac{H_{0}}{\Omega} = \frac{1}{4\lambda^{2}}\alpha^{2} + H_{a},
\end{equation}
where $H_{a} = D + h$ with $D = {\alpha}d$ and $\alpha=2\lambda\beta/\sqrt{\eta}$. The eigenstates of the unperturbed Hamiltonian $h$ are given by $\left|\varepsilon_{i}\right\rangle$ with corresponding eigenvalues $\varepsilon_{1} = -1$, $\varepsilon_{2} = 0$, and $\varepsilon_{3} = 1$. The term $D$ serves as the perturbation. In the limit $\eta \rightarrow \infty$, $\alpha$ can be treated as a perturbation parameter. So we perform perturbation expansion up to the order of $\alpha^{6}$. The ground-state wave function can be expanded as follows:
\begin{equation}
	\begin{aligned}
		|\psi\rangle & = |\varepsilon_{1}\rangle + \sum_{m \ne 1} \frac{|\varepsilon_{m}\rangle\langle \varepsilon_{m}|}{E-\varepsilon_{m}}D|\psi\rangle \\
		& = |\varepsilon_{1}\rangle + G(E)D |\psi\rangle ,
	\end{aligned}
\end{equation}
where $G(E) = {\textstyle \sum_{m\ne 1}} |\varepsilon_{m}\rangle\langle \varepsilon_{m}| / (E-\varepsilon_{m})$ and $H_{a}|\psi\rangle = E|\psi\rangle$. This means that the wave function can be determined through iteration as:
\begin{equation}
	\begin{aligned}
		|\psi\rangle & =  |\varepsilon_{1}\rangle + G(E)D |\varepsilon_{1}\rangle  +  G(E)DG(E)D |\varepsilon_{1}\rangle \\
		& + G(E)DG(E)DG(E)D|\varepsilon_{1}\rangle +  \cdots ,
	\end{aligned}
\end{equation}

Using the relation $H_{a}|\psi\rangle = (D + h)|\psi\rangle = E |\psi\rangle$, we can obtain $D|\psi\rangle = (E-\varepsilon_{1})|\psi\rangle$. It yields the ground-state energy
\begin{equation}
	E - \varepsilon_{1} = \langle \psi|D|\psi\rangle.
\end{equation}
By substituting the wave function into the above equation, we obtain the ground-state energy
\begin{equation}
	\begin{aligned}
		E &= \varepsilon_{1} + \langle \varepsilon_{1}|D|\varepsilon_{1}\rangle + \langle \varepsilon_{1}|DG(E)D|\varepsilon_{1}\rangle \\
		& +  \langle \varepsilon_{1}|DG(E)DG(E)D|\varepsilon_{1}\rangle + \cdots .
	\end{aligned}
	\end{equation}
Clearly, the zero-th energy correction is $E^{(0)} = \varepsilon_{1}$. Moreover, due to the symmetry of the Hamiltonian, the first-order correction is $ \langle \varepsilon_{1}|D|\varepsilon_{1}\rangle = 0$. The second-order correction can be calculated as follows:
\begin{equation}
	\begin{aligned}
		E^{(2)} & =\varepsilon_{1}+\left\langle\varepsilon_{1}|D G(E) D| \varepsilon_{1}\right\rangle \\
		& =\varepsilon_{1}+\frac{\left|D_{12}\right|^{2}}{E^{(0)}-\varepsilon_{2}}=-1-\alpha^{2} \gamma^{2} .
	\end{aligned}
\end{equation}
Similarly, the fourth-order correction of the ground-state energy is obtained as 
\begin{equation}
	\begin{aligned}
		E^{(4)} & =\varepsilon_{1}+\left\langle\varepsilon_{1}|D G(E) D| \varepsilon_{1}\right\rangle \\
		& +\left\langle\varepsilon_{1}|D G(E) DG(E) DG(E) D| \varepsilon_{1}\right\rangle \\
		& =\varepsilon_{1} + \alpha^{2}\frac{\left|D_{12}\right|^{2}}{E^{(2)}-\varepsilon_{2}}  \\
		& + \alpha^{4}\sum_{m \ne 1}\sum_{n \ne 1}\sum_{k \ne 1} \langle \varepsilon_{1}|d \frac{|\varepsilon_m\rangle \langle \varepsilon_m|}{E_{0} - \varepsilon_{m}} d \frac{|\varepsilon_n\rangle \langle \varepsilon_n|}{E_{0} - \varepsilon_{n}} d \frac{|\varepsilon_k\rangle \langle \varepsilon_k|}{E_{0} - \varepsilon_{k}} d |\varepsilon_{1}\rangle \\
		& =-1+\frac{\alpha^{2} \gamma^{2}}{-1 - \alpha^{2} \gamma^{2}} - \frac{1}{2} \alpha^{4} \gamma^{2} .
	\end{aligned}
\end{equation} 
Since $\alpha$ is a small value, the second term of the above energy is approximated as $\alpha^{2} \gamma^{2}(-1+\alpha^{2} \gamma^{2})$. It leads to the approximated ground-state energy 
\begin{equation}
	E^{(4)} = -1 - \alpha^{2} \gamma^{2} + \gamma^{2}(\gamma^{2} - \frac{1}{2}) \alpha^{4}.
\end{equation}

Furthermore, the  ground-state energy can be given up to the sixth-order correction 
\begin{equation}
	\begin{aligned}
		E^{(6)} = & \varepsilon _{1} + \langle \varepsilon_{1}|DG(E)D|\varepsilon_{1}\rangle \\
			& + \langle \varepsilon_{1}|DG(E)DG(E)DG(E)D|\varepsilon_{1}\rangle \\ 
			& + \langle \varepsilon_{1}|DG(E)DG(E)DG(E)DG(E)DG(E)D|\varepsilon_{1}\rangle \\
			= & \varepsilon _{1} + \alpha^{2}\frac{|d_{12}|^{2}}{E^{(4)}-\varepsilon_{2}} \\
			& + \alpha^{4} \sum_{m \ne 1} \sum_{n \ne 1} \sum_{k \ne 1} \langle \varepsilon_{1}|d \frac{|\varepsilon_{m}\rangle \langle \varepsilon_{m}|}{E^{(2)}-\varepsilon _{m}}d\frac{|\varepsilon_n\rangle \langle \varepsilon_n|}{E^{(2)}-\varepsilon _{n}} \\
			& \times d\frac{|\varepsilon_k\rangle \langle \varepsilon_k|}{E^{(2)}-\varepsilon _{k}}d|\varepsilon_{1}\rangle \\
			& + \alpha^{6} \sum_{m \ne 1} \sum_{n \ne 1} \sum_{k \ne 1}\sum_{i \ne 1}\sum_{j \ne 1} \langle \varepsilon_{1}|d \frac{|\varepsilon_m\rangle \langle \varepsilon_m|}{E^{(0)}-\varepsilon _{m}}d\frac{|\varepsilon_n\rangle \langle \varepsilon_n|}{E^{(0)}-\varepsilon _{n}} \\
			& \times d\frac{|\varepsilon_k\rangle \langle \varepsilon_k|}{E^{(0)}-\varepsilon _{k}}d\frac{|\varepsilon_i\rangle \langle \varepsilon_i|}{E^{(0)}-\varepsilon _{i}}d\frac{|\varepsilon_j\rangle \langle \varepsilon_j|}{E^{(0)}-\varepsilon _{j}}d|\varepsilon_{1}\rangle\\
			= & \varepsilon_{1} + \frac{\alpha^{2}\gamma^{2}}{E^{(4)} - \varepsilon_{2}} + \alpha^{4}\frac{|d_{12}|^{2}}{(E^{(2)} - \varepsilon_{2})^{2}}\frac{|d_{23}|^{2}}{(E^{(2)} - \varepsilon_{3})^{2}} \\
			& + \alpha^{6}\frac{|d_{12}|^{3}}{(E^{(0)} - \varepsilon_{2})^{3}}\frac{|d_{23}|^{3}}{(E^{(0)} - \varepsilon_{2})^{3}} .
	\end{aligned}
\end{equation}
With the sixth-order correction, the ground-state energy of the effective Hamiltonian (\ref{effham}) can be approximately given up to the order of $\alpha^{6}$
\begin{equation}
	\begin{aligned}
		E_{SR}/\Omega = & \frac{1}{4\lambda^{2}}\alpha^{2} -1 - \alpha^{2} \gamma^{2} + \gamma^{2}(\gamma^{2} - \frac{1}{2}) \alpha^{4} \\
		& - \frac{1}{4} \gamma^{2}(8\gamma^{4} - 7\gamma^{2} +1) \alpha^{6}\\
  &= c_{1} \alpha^{2} + c_{2}\alpha^{4} + c_{3} \alpha^{6}-1,\\
	\end{aligned}
\end{equation}
where the coefficients are $c_{1} = 1/(4 \lambda^{2}) - \gamma^{2}$,~$c_{2} = \gamma^{2}(\gamma^{2} - \frac{1}{2})$, and~$c_{3} = -\gamma^{2}(1-7\gamma^{2} + 8\gamma^{4})/4$ . The energy is given in Eq.(\ref{mfenergy}).

\begin{figure}[htbp]
	\centering
	\includegraphics[width=0.45\textwidth]{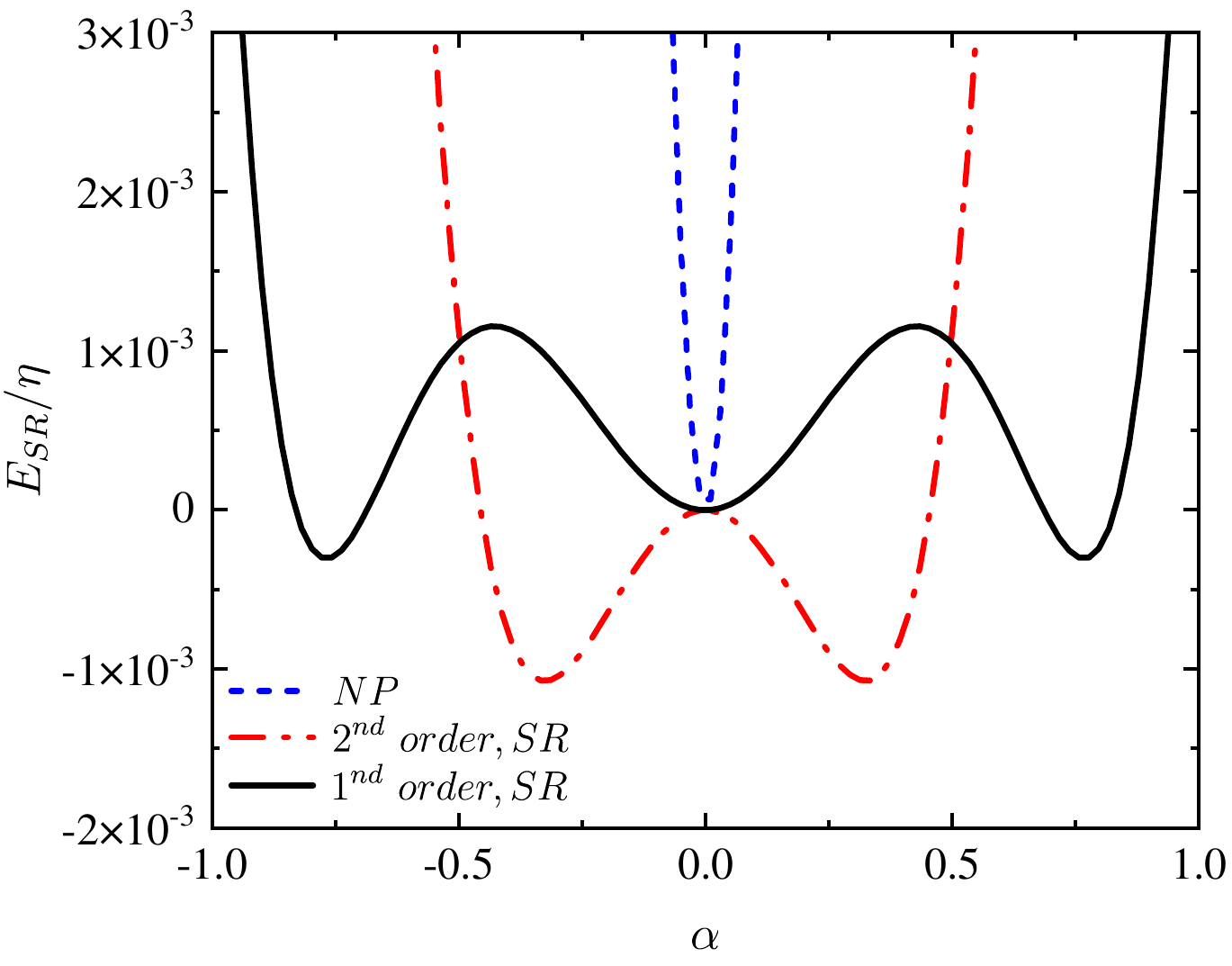}
	\caption{Ground-state energy~$E_{SR}/\eta$ as a function of the order parameter~$\alpha$ in the superradiant phases of the first-order QPT ($\gamma = 0.6 < \gamma_{\texttt{TCP}}, \lambda = 0.82>\lambda_c$) in black solid line, and the second-order QCP ($\gamma = 0.8 > \gamma_{\texttt{TCP}}, \lambda = 0.65>\lambda_c$) in red dashed line with $\omega=1$. The energy in the normal phase is listed for $\gamma = 0.1 < \gamma_{\texttt{TCP}}, \lambda = 0.5<\lambda_c$~~(blue dotted line).} 
	\label{Energy}
\end{figure}

Fig.\ref{Energy} depicts the ground-state energy $E_{SR}/\eta$ as a function of $\alpha$ for various values of $\lambda$ and $\gamma$. By adjusting the transition ratio $\gamma<\gamma_{\texttt{TCP}} $, $E_{SR}$ locates at three minimal value of $\alpha$, predicting a first-order phase transition.

%\nocite{*}
\bibliography{apssamp}% Produces the bibliography via BibTeX.

\begin{thebibliography}{44}%
\makeatletter
\providecommand \@ifxundefined [1]{%
 \@ifx{#1\undefined}
}%
\providecommand \@ifnum [1]{%
 \ifnum #1\expandafter \@firstoftwo
 \else \expandafter \@secondoftwo
 \fi
}%
\providecommand \@ifx [1]{%
 \ifx #1\expandafter \@firstoftwo
 \else \expandafter \@secondoftwo
 \fi
}%
\providecommand \natexlab [1]{#1}%
\providecommand \enquote  [1]{``#1''}%
\providecommand \bibnamefont  [1]{#1}%
\providecommand \bibfnamefont [1]{#1}%
\providecommand \citenamefont [1]{#1}%
\providecommand \href@noop [0]{\@secondoftwo}%
\providecommand \href [0]{\begingroup \@sanitize@url \@href}%
\providecommand \@href[1]{\@@startlink{#1}\@@href}%
\providecommand \@@href[1]{\endgroup#1\@@endlink}%
\providecommand \@sanitize@url [0]{\catcode `\\12\catcode `\$12\catcode
  `\&12\catcode `\#12\catcode `\^12\catcode `\_12\catcode `\%12\relax}%
\providecommand \@@startlink[1]{}%
\providecommand \@@endlink[0]{}%
\providecommand \url  [0]{\begingroup\@sanitize@url \@url }%
\providecommand \@url [1]{\endgroup\@href {#1}{\urlprefix }}%
\providecommand \urlprefix  [0]{URL }%
\providecommand \Eprint [0]{\href }%
\providecommand \doibase [0]{https://doi.org/}%
\providecommand \selectlanguage [0]{\@gobble}%
\providecommand \bibinfo  [0]{\@secondoftwo}%
\providecommand \bibfield  [0]{\@secondoftwo}%
\providecommand \translation [1]{[#1]}%
\providecommand \BibitemOpen [0]{}%
\providecommand \bibitemStop [0]{}%
\providecommand \bibitemNoStop [0]{.\EOS\space}%
\providecommand \EOS [0]{\spacefactor3000\relax}%
\providecommand \BibitemShut  [1]{\csname bibitem#1\endcsname}%
\let\auto@bib@innerbib\@empty
%</preamble>
\bibitem [{\citenamefont {Sachdev}(2011)}]{Sachdev}%
  \BibitemOpen
  \bibfield  {author} {\bibinfo {author} {\bibfnamefont {S.}~\bibnamefont
  {Sachdev}},\ }\href {https://doi.org/10.1017/CBO9780511973765} {\emph
  {\bibinfo {title} {Quantum Phase Transitions}}},\ \bibinfo {edition} {2nd}\
  ed.\ (\bibinfo  {publisher} {Cambridge University Press},\ \bibinfo {address}
  {Cambridge},\ \bibinfo {year} {2011})\BibitemShut {NoStop}%
\bibitem [{\citenamefont {L\"ohneysen}\ \emph {et~al.}(2007)\citenamefont
  {L\"ohneysen}, \citenamefont {Rosch}, \citenamefont {Vojta},\ and\
  \citenamefont {W\"olfle}}]{RevModPhys2007}%
  \BibitemOpen
  \bibfield  {author} {\bibinfo {author} {\bibfnamefont {H.~v.}\ \bibnamefont
  {L\"ohneysen}}, \bibinfo {author} {\bibfnamefont {A.}~\bibnamefont {Rosch}},
  \bibinfo {author} {\bibfnamefont {M.}~\bibnamefont {Vojta}},\ and\ \bibinfo
  {author} {\bibfnamefont {P.}~\bibnamefont {W\"olfle}},\ }\bibfield  {title}
  {\bibinfo {title} {Fermi-liquid instabilities at magnetic quantum phase
  transitions},\ }\href {https://doi.org/10.1103/RevModPhys.79.1015} {\bibfield
   {journal} {\bibinfo  {journal} {Rev. Mod. Phys.}\ }\textbf {\bibinfo
  {volume} {79}},\ \bibinfo {pages} {1015} (\bibinfo {year}
  {2007})}\BibitemShut {NoStop}%
\bibitem [{\citenamefont {Sondhi}\ \emph {et~al.}(1997)\citenamefont {Sondhi},
  \citenamefont {Girvin}, \citenamefont {Carini},\ and\ \citenamefont
  {Shahar}}]{RevModPhys.69.315}%
  \BibitemOpen
  \bibfield  {author} {\bibinfo {author} {\bibfnamefont {S.~L.}\ \bibnamefont
  {Sondhi}}, \bibinfo {author} {\bibfnamefont {S.~M.}\ \bibnamefont {Girvin}},
  \bibinfo {author} {\bibfnamefont {J.~P.}\ \bibnamefont {Carini}},\ and\
  \bibinfo {author} {\bibfnamefont {D.}~\bibnamefont {Shahar}},\ }\bibfield
  {title} {\bibinfo {title} {Continuous quantum phase transitions},\ }\href
  {https://doi.org/10.1103/RevModPhys.69.315} {\bibfield  {journal} {\bibinfo
  {journal} {Rev. Mod. Phys.}\ }\textbf {\bibinfo {volume} {69}},\ \bibinfo
  {pages} {315} (\bibinfo {year} {1997})}\BibitemShut {NoStop}%
\bibitem [{\citenamefont {Griffiths}(1975)}]{griff}%
  \BibitemOpen
  \bibfield  {author} {\bibinfo {author} {\bibfnamefont {R.~B.}\ \bibnamefont
  {Griffiths}},\ }\bibfield  {title} {\bibinfo {title} {Phase diagrams and
  higher-order critical points},\ }\href@noop {} {\bibfield  {journal}
  {\bibinfo  {journal} {Phys. Rev. B}\ }\textbf {\bibinfo {volume} {12}},\
  \bibinfo {pages} {345} (\bibinfo {year} {1975})}\BibitemShut {NoStop}%
\bibitem [{\citenamefont {Belitz}\ \emph {et~al.}(1999)\citenamefont {Belitz},
  \citenamefont {Kirkpatrick},\ and\ \citenamefont {Vojta}}]{PhysRevLett1999}%
  \BibitemOpen
  \bibfield  {author} {\bibinfo {author} {\bibfnamefont {D.}~\bibnamefont
  {Belitz}}, \bibinfo {author} {\bibfnamefont {T.~R.}\ \bibnamefont
  {Kirkpatrick}},\ and\ \bibinfo {author} {\bibfnamefont {T.}~\bibnamefont
  {Vojta}},\ }\bibfield  {title} {\bibinfo {title} {First order transitions and
  multicritical points in weak itinerant ferromagnets},\ }\href
  {https://doi.org/10.1103/PhysRevLett.82.4707} {\bibfield  {journal} {\bibinfo
   {journal} {Phys. Rev. Lett.}\ }\textbf {\bibinfo {volume} {82}},\ \bibinfo
  {pages} {4707} (\bibinfo {year} {1999})}\BibitemShut {NoStop}%
\bibitem [{\citenamefont {Belitz}\ and\ \citenamefont
  {Kirkpatrick}(2017)}]{Belitz}%
  \BibitemOpen
  \bibfield  {author} {\bibinfo {author} {\bibfnamefont {D.}~\bibnamefont
  {Belitz}}\ and\ \bibinfo {author} {\bibfnamefont {T.~R.}\ \bibnamefont
  {Kirkpatrick}},\ }\bibfield  {title} {\bibinfo {title} {Quantum triple point
  and quantum critical end points in metallic magnets},\ }\href@noop {}
  {\bibfield  {journal} {\bibinfo  {journal} {Phys. Rev. Lett.}\ }\textbf
  {\bibinfo {volume} {119}},\ \bibinfo {pages} {267202} (\bibinfo {year}
  {2017})}\BibitemShut {NoStop}%
\bibitem [{\citenamefont {Kaluarachchi}\ \emph {et~al.}(2018)\citenamefont
  {Kaluarachchi}, \citenamefont {Taufour}, \citenamefont {Bud'ko},\ and\
  \citenamefont {Canfield}}]{Canfield}%
  \BibitemOpen
  \bibfield  {author} {\bibinfo {author} {\bibfnamefont {U.~S.}\ \bibnamefont
  {Kaluarachchi}}, \bibinfo {author} {\bibfnamefont {V.}~\bibnamefont
  {Taufour}}, \bibinfo {author} {\bibfnamefont {S.~L.}\ \bibnamefont
  {Bud'ko}},\ and\ \bibinfo {author} {\bibfnamefont {P.~C.}\ \bibnamefont
  {Canfield}},\ }\bibfield  {title} {\bibinfo {title} {Quantum tricritical
  point in the temperature-pressure-magnetic field phase diagram of
  ${\mathrm{cetige}}_{3}$},\ }\href@noop {} {\bibfield  {journal} {\bibinfo
  {journal} {Phys. Rev. B}\ }\textbf {\bibinfo {volume} {97}},\ \bibinfo
  {pages} {045139} (\bibinfo {year} {2018})}\BibitemShut {NoStop}%
\bibitem [{\citenamefont {Wu}\ \emph {et~al.}(2019)\citenamefont {Wu},
  \citenamefont {Guo}, \citenamefont {Chen}, \citenamefont {Su}, \citenamefont
  {Wang}, \citenamefont {Smidman},\ and\ \citenamefont {Yuan}}]{yuan2019}%
  \BibitemOpen
  \bibfield  {author} {\bibinfo {author} {\bibfnamefont {F.}~\bibnamefont
  {Wu}}, \bibinfo {author} {\bibfnamefont {C.~Y.}\ \bibnamefont {Guo}},
  \bibinfo {author} {\bibfnamefont {Y.}~\bibnamefont {Chen}}, \bibinfo {author}
  {\bibfnamefont {H.}~\bibnamefont {Su}}, \bibinfo {author} {\bibfnamefont
  {A.}~\bibnamefont {Wang}}, \bibinfo {author} {\bibfnamefont {M.}~\bibnamefont
  {Smidman}},\ and\ \bibinfo {author} {\bibfnamefont {H.~Q.}\ \bibnamefont
  {Yuan}},\ }\bibfield  {title} {\bibinfo {title} {Magnetic field induced
  antiferromagnetic tricritical points in $\mathrm{Ce}{}_{2}\mathrm{Sb}$ and
  $\mathrm{Ce}{}_{2}\mathrm{Bi}$},\ }\href@noop {} {\bibfield  {journal}
  {\bibinfo  {journal} {Phys. Rev. B}\ }\textbf {\bibinfo {volume} {99}},\
  \bibinfo {pages} {064419} (\bibinfo {year} {2019})}\BibitemShut {NoStop}%
\bibitem [{\citenamefont {Friedemann}\ \emph {et~al.}(2017)\citenamefont
  {Friedemann}, \citenamefont {Duncan}, \citenamefont {Hirschberger},
  \citenamefont {Bauer}, \citenamefont {Kuchler}, \citenamefont {Neubauer},
  \citenamefont {Brando}, \citenamefont {Pfleiderer},\ and\ \citenamefont
  {Grosche}}]{Friedemann}%
  \BibitemOpen
  \bibfield  {author} {\bibinfo {author} {\bibfnamefont {S.}~\bibnamefont
  {Friedemann}}, \bibinfo {author} {\bibfnamefont {W.~J.}\ \bibnamefont
  {Duncan}}, \bibinfo {author} {\bibfnamefont {M.}~\bibnamefont
  {Hirschberger}}, \bibinfo {author} {\bibfnamefont {T.}~\bibnamefont {Bauer}},
  \bibinfo {author} {\bibfnamefont {R.}~\bibnamefont {Kuchler}}, \bibinfo
  {author} {\bibfnamefont {A.}~\bibnamefont {Neubauer}}, \bibinfo {author}
  {\bibfnamefont {M.}~\bibnamefont {Brando}}, \bibinfo {author} {\bibfnamefont
  {C.}~\bibnamefont {Pfleiderer}},\ and\ \bibinfo {author} {\bibfnamefont
  {F.~M.}\ \bibnamefont {Grosche}},\ }\bibfield  {title} {\bibinfo {title}
  {Quantum tricritical points in nbfe2},\ }\href@noop {} {\bibfield  {journal}
  {\bibinfo  {journal} {Nature Physics}\ }\textbf {\bibinfo {volume} {14}},\
  \bibinfo {pages} {62} (\bibinfo {year} {2017})}\BibitemShut {NoStop}%
\bibitem [{\citenamefont {Xu}\ and\ \citenamefont {Pu}(2019)}]{pu}%
  \BibitemOpen
  \bibfield  {author} {\bibinfo {author} {\bibfnamefont {Y.}~\bibnamefont
  {Xu}}\ and\ \bibinfo {author} {\bibfnamefont {H.}~\bibnamefont {Pu}},\
  }\bibfield  {title} {\bibinfo {title} {Emergent universality in a quantum
  tricritical dicke model},\ }\href@noop {} {\bibfield  {journal} {\bibinfo
  {journal} {Phys. Rev. Lett.}\ }\textbf {\bibinfo {volume} {122}},\ \bibinfo
  {pages} {193201} (\bibinfo {year} {2019})}\BibitemShut {NoStop}%
\bibitem [{\citenamefont {Yin}\ \emph {et~al.}(2018)\citenamefont {Yin},
  \citenamefont {Jian},\ and\ \citenamefont {Yao}}]{PhysRevLett2018}%
  \BibitemOpen
  \bibfield  {author} {\bibinfo {author} {\bibfnamefont {S.}~\bibnamefont
  {Yin}}, \bibinfo {author} {\bibfnamefont {S.-K.}\ \bibnamefont {Jian}},\ and\
  \bibinfo {author} {\bibfnamefont {H.}~\bibnamefont {Yao}},\ }\bibfield
  {title} {\bibinfo {title} {Chiral tricritical point: A new universality class
  in dirac systems},\ }\href {https://doi.org/10.1103/PhysRevLett.120.215702}
  {\bibfield  {journal} {\bibinfo  {journal} {Phys. Rev. Lett.}\ }\textbf
  {\bibinfo {volume} {120}},\ \bibinfo {pages} {215702} (\bibinfo {year}
  {2018})}\BibitemShut {NoStop}%
\bibitem [{\citenamefont {Kato}\ and\ \citenamefont
  {Misawa}(2015)}]{PhysRevB2015}%
  \BibitemOpen
  \bibfield  {author} {\bibinfo {author} {\bibfnamefont {Y.}~\bibnamefont
  {Kato}}\ and\ \bibinfo {author} {\bibfnamefont {T.}~\bibnamefont {Misawa}},\
  }\bibfield  {title} {\bibinfo {title} {Quantum tricriticality in
  antiferromagnetic ising model with transverse field: A quantum monte carlo
  study},\ }\href {https://doi.org/10.1103/PhysRevB.92.174419} {\bibfield
  {journal} {\bibinfo  {journal} {Phys. Rev. B}\ }\textbf {\bibinfo {volume}
  {92}},\ \bibinfo {pages} {174419} (\bibinfo {year} {2015})}\BibitemShut
  {NoStop}%
\bibitem [{\citenamefont {Cheng}\ \emph {et~al.}(2022)\citenamefont {Cheng},
  \citenamefont {Fallas~Padilla}, \citenamefont {Deng}, \citenamefont {Zhang},\
  and\ \citenamefont {Pu}}]{zhang2022}%
  \BibitemOpen
  \bibfield  {author} {\bibinfo {author} {\bibfnamefont {G.-J.}\ \bibnamefont
  {Cheng}}, \bibinfo {author} {\bibfnamefont {D.}~\bibnamefont
  {Fallas~Padilla}}, \bibinfo {author} {\bibfnamefont {T.}~\bibnamefont
  {Deng}}, \bibinfo {author} {\bibfnamefont {Y.-Y.}\ \bibnamefont {Zhang}},\
  and\ \bibinfo {author} {\bibfnamefont {H.}~\bibnamefont {Pu}},\ }\bibfield
  {title} {\bibinfo {title} {Chiral quantum phases and tricriticality in a
  dicke triangle},\ }\href {https://doi.org/10.1007/s44214-022-00019-5}
  {\bibfield  {journal} {\bibinfo  {journal} {Quantum Frontiers}\ }\textbf
  {\bibinfo {volume} {1}},\ \bibinfo {pages} {2731} (\bibinfo {year}
  {2022})}\BibitemShut {NoStop}%
\bibitem [{\citenamefont {Fallas~Padilla}\ and\ \citenamefont
  {Pu}(2023)}]{PhysRevA.108.033706}%
  \BibitemOpen
  \bibfield  {author} {\bibinfo {author} {\bibfnamefont {D.}~\bibnamefont
  {Fallas~Padilla}}\ and\ \bibinfo {author} {\bibfnamefont {H.}~\bibnamefont
  {Pu}},\ }\bibfield  {title} {\bibinfo {title} {Tricritical dicke model with
  and without dissipation},\ }\href
  {https://doi.org/10.1103/PhysRevA.108.033706} {\bibfield  {journal} {\bibinfo
   {journal} {Phys. Rev. A}\ }\textbf {\bibinfo {volume} {108}},\ \bibinfo
  {pages} {033706} (\bibinfo {year} {2023})}\BibitemShut {NoStop}%
\bibitem [{\citenamefont {Greentree}\ \emph {et~al.}(2006)\citenamefont
  {Greentree}, \citenamefont {Tahan}, \citenamefont {Cole},\ and\ \citenamefont
  {Hollenberg}}]{Greentree2006}%
  \BibitemOpen
  \bibfield  {author} {\bibinfo {author} {\bibfnamefont {A.~D.}\ \bibnamefont
  {Greentree}}, \bibinfo {author} {\bibfnamefont {C.}~\bibnamefont {Tahan}},
  \bibinfo {author} {\bibfnamefont {J.~H.}\ \bibnamefont {Cole}},\ and\
  \bibinfo {author} {\bibfnamefont {L.~C.}\ \bibnamefont {Hollenberg}},\
  }\bibfield  {title} {\bibinfo {title} {Quantum phase transitions of light},\
  }\href@noop {} {\bibfield  {journal} {\bibinfo  {journal} {Nature Physics}\
  }\textbf {\bibinfo {volume} {2}},\ \bibinfo {pages} {856} (\bibinfo {year}
  {2006})}\BibitemShut {NoStop}%
\bibitem [{\citenamefont {Hartmann}\ \emph {et~al.}(2006)\citenamefont
  {Hartmann}, \citenamefont {Brandao},\ and\ \citenamefont {Plenio}}]{plenio}%
  \BibitemOpen
  \bibfield  {author} {\bibinfo {author} {\bibfnamefont {M.~J.}\ \bibnamefont
  {Hartmann}}, \bibinfo {author} {\bibfnamefont {F.~G.}\ \bibnamefont
  {Brandao}},\ and\ \bibinfo {author} {\bibfnamefont {M.~B.}\ \bibnamefont
  {Plenio}},\ }\bibfield  {title} {\bibinfo {title} {Strongly interacting
  polaritons in coupled arrays of cavities},\ }\href@noop {} {\bibfield
  {journal} {\bibinfo  {journal} {Nature Physics}\ }\textbf {\bibinfo {volume}
  {2}},\ \bibinfo {pages} {849} (\bibinfo {year} {2006})}\BibitemShut {NoStop}%
\bibitem [{\citenamefont {Zhu}\ \emph {et~al.}(2020)\citenamefont {Zhu},
  \citenamefont {Ping}, \citenamefont {Yang},\ and\ \citenamefont
  {Agarwal}}]{zhu2020}%
  \BibitemOpen
  \bibfield  {author} {\bibinfo {author} {\bibfnamefont {C.}~\bibnamefont
  {Zhu}}, \bibinfo {author} {\bibfnamefont {L.}~\bibnamefont {Ping}}, \bibinfo
  {author} {\bibfnamefont {Y.}~\bibnamefont {Yang}},\ and\ \bibinfo {author}
  {\bibfnamefont {G.~S.}\ \bibnamefont {Agarwal}},\ }\bibfield  {title}
  {\bibinfo {title} {Squeezed light induced symmetry breaking superradiant
  phase transition},\ }\href@noop {} {\bibfield  {journal} {\bibinfo  {journal}
  {Physical Review Letters}\ }\textbf {\bibinfo {volume} {124}},\ \bibinfo
  {pages} {073602} (\bibinfo {year} {2020})}\BibitemShut {NoStop}%
\bibitem [{\citenamefont {Dicke}(1954)}]{Dicke}%
  \BibitemOpen
  \bibfield  {author} {\bibinfo {author} {\bibfnamefont {R.~H.}\ \bibnamefont
  {Dicke}},\ }\bibfield  {title} {\bibinfo {title} {Coherence in spontaneous
  radiation processes},\ }\href@noop {} {\bibfield  {journal} {\bibinfo
  {journal} {Phys. Rev.}\ }\textbf {\bibinfo {volume} {93}},\ \bibinfo {pages}
  {99} (\bibinfo {year} {1954})}\BibitemShut {NoStop}%
\bibitem [{\citenamefont {Lambert}\ \emph {et~al.}(2004)\citenamefont
  {Lambert}, \citenamefont {Emary},\ and\ \citenamefont {Brandes}}]{lambert}%
  \BibitemOpen
  \bibfield  {author} {\bibinfo {author} {\bibfnamefont {N.}~\bibnamefont
  {Lambert}}, \bibinfo {author} {\bibfnamefont {C.}~\bibnamefont {Emary}},\
  and\ \bibinfo {author} {\bibfnamefont {T.}~\bibnamefont {Brandes}},\
  }\bibfield  {title} {\bibinfo {title} {Entanglement and the phase transition
  in single-mode superradiance},\ }\href@noop {} {\bibfield  {journal}
  {\bibinfo  {journal} {Phys. Rev. Lett.}\ }\textbf {\bibinfo {volume} {92}},\
  \bibinfo {pages} {073602} (\bibinfo {year} {2004})}\BibitemShut {NoStop}%
\bibitem [{\citenamefont {Chen}\ \emph {et~al.}(2008)\citenamefont {Chen},
  \citenamefont {Zhang}, \citenamefont {Liu},\ and\ \citenamefont
  {Wang}}]{chen2008}%
  \BibitemOpen
  \bibfield  {author} {\bibinfo {author} {\bibfnamefont {Q.-H.}\ \bibnamefont
  {Chen}}, \bibinfo {author} {\bibfnamefont {Y.-Y.}\ \bibnamefont {Zhang}},
  \bibinfo {author} {\bibfnamefont {T.}~\bibnamefont {Liu}},\ and\ \bibinfo
  {author} {\bibfnamefont {K.-L.}\ \bibnamefont {Wang}},\ }\bibfield  {title}
  {\bibinfo {title} {Numerically exact solution to the finite-size dicke
  model},\ }\href@noop {} {\bibfield  {journal} {\bibinfo  {journal} {Phys.
  Rev. A}\ }\textbf {\bibinfo {volume} {78}},\ \bibinfo {pages} {051801}
  (\bibinfo {year} {2008})}\BibitemShut {NoStop}%
\bibitem [{\citenamefont {Chen}\ and\ \citenamefont {Zhang}(2018)}]{zhang2019}%
  \BibitemOpen
  \bibfield  {author} {\bibinfo {author} {\bibfnamefont {X.-Y.}\ \bibnamefont
  {Chen}}\ and\ \bibinfo {author} {\bibfnamefont {Y.-Y.}\ \bibnamefont
  {Zhang}},\ }\bibfield  {title} {\bibinfo {title} {Finite-size scaling
  analysis in the two-photon dicke model},\ }\href@noop {} {\bibfield
  {journal} {\bibinfo  {journal} {Phys. Rev. A}\ }\textbf {\bibinfo {volume}
  {97}},\ \bibinfo {pages} {053821} (\bibinfo {year} {2018})}\BibitemShut
  {NoStop}%
\bibitem [{\citenamefont {Emary}\ and\ \citenamefont
  {Brandes}(2003)}]{Emary03}%
  \BibitemOpen
  \bibfield  {author} {\bibinfo {author} {\bibfnamefont {C.}~\bibnamefont
  {Emary}}\ and\ \bibinfo {author} {\bibfnamefont {T.}~\bibnamefont
  {Brandes}},\ }\bibfield  {title} {\bibinfo {title} {Chaos and the quantum
  phase transition in the dicke model},\ }\href
  {https://doi.org/10.1103/PhysRevE.67.066203} {\bibfield  {journal} {\bibinfo
  {journal} {Phys. Rev. E}\ }\textbf {\bibinfo {volume} {67}},\ \bibinfo
  {pages} {066203} (\bibinfo {year} {2003})}\BibitemShut {NoStop}%
\bibitem [{\citenamefont {Nagy}\ \emph {et~al.}(2010)\citenamefont {Nagy},
  \citenamefont {K\'onya}, \citenamefont {Szirmai},\ and\ \citenamefont
  {Domokos}}]{nagy}%
  \BibitemOpen
  \bibfield  {author} {\bibinfo {author} {\bibfnamefont {D.}~\bibnamefont
  {Nagy}}, \bibinfo {author} {\bibfnamefont {G.}~\bibnamefont {K\'onya}},
  \bibinfo {author} {\bibfnamefont {G.}~\bibnamefont {Szirmai}},\ and\ \bibinfo
  {author} {\bibfnamefont {P.}~\bibnamefont {Domokos}},\ }\bibfield  {title}
  {\bibinfo {title} {Dicke-model phase transition in the quantum motion of a
  bose-einstein condensate in an optical cavity},\ }\href@noop {} {\bibfield
  {journal} {\bibinfo  {journal} {Phys. Rev. Lett.}\ }\textbf {\bibinfo
  {volume} {104}},\ \bibinfo {pages} {130401} (\bibinfo {year}
  {2010})}\BibitemShut {NoStop}%
\bibitem [{\citenamefont {zhang}\ \emph {et~al.}(2021)\citenamefont {zhang},
  \citenamefont {Yu}, \citenamefont {Zemao}, \citenamefont {Juan},
  \citenamefont {Jijie}, \citenamefont {Shujin},\ and\ \citenamefont
  {Haibin}}]{sciencewu}%
  \BibitemOpen
  \bibfield  {author} {\bibinfo {author} {\bibfnamefont {X.}~\bibnamefont
  {zhang}}, \bibinfo {author} {\bibfnamefont {C.}~\bibnamefont {Yu}}, \bibinfo
  {author} {\bibfnamefont {W.}~\bibnamefont {Zemao}}, \bibinfo {author}
  {\bibfnamefont {W.}~\bibnamefont {Juan}}, \bibinfo {author} {\bibfnamefont
  {F.}~\bibnamefont {Jijie}}, \bibinfo {author} {\bibfnamefont
  {D.}~\bibnamefont {Shujin}},\ and\ \bibinfo {author} {\bibfnamefont
  {W.}~\bibnamefont {Haibin}},\ }\bibfield  {title} {\bibinfo {title}
  {Observation of a superradiant quantum phase transition in an intracavity
  degenerate fermi gas},\ }\href {https://doi.org/10.1126/science.abd4385}
  {\bibfield  {journal} {\bibinfo  {journal} {Science}\ }\textbf {\bibinfo
  {volume} {373}},\ \bibinfo {pages} {1359} (\bibinfo {year}
  {2021})}\BibitemShut {NoStop}%
\bibitem [{\citenamefont {Ashhab}(2013)}]{Ashhab2013}%
  \BibitemOpen
  \bibfield  {author} {\bibinfo {author} {\bibfnamefont {S.}~\bibnamefont
  {Ashhab}},\ }\bibfield  {title} {\bibinfo {title} {Superradiance transition
  in a system with a single qubit and a single oscillator},\ }\href@noop {}
  {\bibfield  {journal} {\bibinfo  {journal} {Phys. Rev. A}\ }\textbf {\bibinfo
  {volume} {87}},\ \bibinfo {pages} {013826} (\bibinfo {year}
  {2013})}\BibitemShut {NoStop}%
\bibitem [{\citenamefont {Hwang}\ \emph {et~al.}(2015)\citenamefont {Hwang},
  \citenamefont {Puebla},\ and\ \citenamefont
  {Plenio}}]{PhysRevLett.115.180404}%
  \BibitemOpen
  \bibfield  {author} {\bibinfo {author} {\bibfnamefont {M.-J.}\ \bibnamefont
  {Hwang}}, \bibinfo {author} {\bibfnamefont {R.}~\bibnamefont {Puebla}},\ and\
  \bibinfo {author} {\bibfnamefont {M.~B.}\ \bibnamefont {Plenio}},\ }\bibfield
   {title} {\bibinfo {title} {Quantum phase transition and universal dynamics
  in the rabi model},\ }\href {https://doi.org/10.1103/PhysRevLett.115.180404}
  {\bibfield  {journal} {\bibinfo  {journal} {Phys. Rev. Lett.}\ }\textbf
  {\bibinfo {volume} {115}},\ \bibinfo {pages} {180404} (\bibinfo {year}
  {2015})}\BibitemShut {NoStop}%
\bibitem [{\citenamefont {Liu}\ \emph {et~al.}(2017)\citenamefont {Liu},
  \citenamefont {Chesi}, \citenamefont {Ying}, \citenamefont {Chen},
  \citenamefont {Luo},\ and\ \citenamefont {Lin}}]{PhysRevLett.119.220601}%
  \BibitemOpen
  \bibfield  {author} {\bibinfo {author} {\bibfnamefont {M.}~\bibnamefont
  {Liu}}, \bibinfo {author} {\bibfnamefont {S.}~\bibnamefont {Chesi}}, \bibinfo
  {author} {\bibfnamefont {Z.-J.}\ \bibnamefont {Ying}}, \bibinfo {author}
  {\bibfnamefont {X.}~\bibnamefont {Chen}}, \bibinfo {author} {\bibfnamefont
  {H.-G.}\ \bibnamefont {Luo}},\ and\ \bibinfo {author} {\bibfnamefont {H.-Q.}\
  \bibnamefont {Lin}},\ }\bibfield  {title} {\bibinfo {title} {Universal
  scaling and critical exponents of the anisotropic quantum rabi model},\
  }\href {https://doi.org/10.1103/PhysRevLett.119.220601} {\bibfield  {journal}
  {\bibinfo  {journal} {Phys. Rev. Lett.}\ }\textbf {\bibinfo {volume} {119}},\
  \bibinfo {pages} {220601} (\bibinfo {year} {2017})}\BibitemShut {NoStop}%
\bibitem [{\citenamefont {Chen}\ \emph
  {et~al.}(2020{\natexlab{a}})\citenamefont {Chen}, \citenamefont {Zhang},
  \citenamefont {Fu},\ and\ \citenamefont {Zheng}}]{PhysRevA.101.033827}%
  \BibitemOpen
  \bibfield  {author} {\bibinfo {author} {\bibfnamefont {X.-Y.}\ \bibnamefont
  {Chen}}, \bibinfo {author} {\bibfnamefont {Y.-Y.}\ \bibnamefont {Zhang}},
  \bibinfo {author} {\bibfnamefont {L.}~\bibnamefont {Fu}},\ and\ \bibinfo
  {author} {\bibfnamefont {H.}~\bibnamefont {Zheng}},\ }\bibfield  {title}
  {\bibinfo {title} {Generalized coherent-squeezed-state expansion for the
  super-radiant phase transition},\ }\href
  {https://doi.org/10.1103/PhysRevA.101.033827} {\bibfield  {journal} {\bibinfo
   {journal} {Phys. Rev. A}\ }\textbf {\bibinfo {volume} {101}},\ \bibinfo
  {pages} {033827} (\bibinfo {year} {2020}{\natexlab{a}})}\BibitemShut
  {NoStop}%
\bibitem [{\citenamefont {L\"u}\ \emph {et~al.}(2018)\citenamefont {L\"u},
  \citenamefont {Zheng}, \citenamefont {Zhu},\ and\ \citenamefont
  {Wu}}]{PhysRevApplied.9.064006}%
  \BibitemOpen
  \bibfield  {author} {\bibinfo {author} {\bibfnamefont {X.-Y.}\ \bibnamefont
  {L\"u}}, \bibinfo {author} {\bibfnamefont {L.-L.}\ \bibnamefont {Zheng}},
  \bibinfo {author} {\bibfnamefont {G.-L.}\ \bibnamefont {Zhu}},\ and\ \bibinfo
  {author} {\bibfnamefont {Y.}~\bibnamefont {Wu}},\ }\bibfield  {title}
  {\bibinfo {title} {Single-photon-triggered quantum phase transition},\ }\href
  {https://doi.org/10.1103/PhysRevApplied.9.064006} {\bibfield  {journal}
  {\bibinfo  {journal} {Phys. Rev. Appl.}\ }\textbf {\bibinfo {volume} {9}},\
  \bibinfo {pages} {064006} (\bibinfo {year} {2018})}\BibitemShut {NoStop}%
\bibitem [{\citenamefont {Chen}\ \emph {et~al.}(2021)\citenamefont {Chen},
  \citenamefont {Wu}, \citenamefont {Jiang}, \citenamefont {L{\"u}},
  \citenamefont {Peng},\ and\ \citenamefont {Du}}]{chen2021}%
  \BibitemOpen
  \bibfield  {author} {\bibinfo {author} {\bibfnamefont {X.}~\bibnamefont
  {Chen}}, \bibinfo {author} {\bibfnamefont {Z.}~\bibnamefont {Wu}}, \bibinfo
  {author} {\bibfnamefont {M.}~\bibnamefont {Jiang}}, \bibinfo {author}
  {\bibfnamefont {X.-Y.}\ \bibnamefont {L{\"u}}}, \bibinfo {author}
  {\bibfnamefont {X.}~\bibnamefont {Peng}},\ and\ \bibinfo {author}
  {\bibfnamefont {J.}~\bibnamefont {Du}},\ }\bibfield  {title} {\bibinfo
  {title} {Experimental quantum simulation of superradiant phase transition
  beyond no-go theorem via antisqueezing},\ }\href@noop {} {\bibfield
  {journal} {\bibinfo  {journal} {Nat. Commun.}\ }\textbf {\bibinfo {volume}
  {12}},\ \bibinfo {pages} {1} (\bibinfo {year} {2021})}\BibitemShut {NoStop}%
\bibitem [{\citenamefont {Cai}\ and\ \citenamefont {et~al.}(2021)}]{NCcai2021}%
  \BibitemOpen
  \bibfield  {author} {\bibinfo {author} {\bibfnamefont {M.~L.}\ \bibnamefont
  {Cai}}\ and\ \bibinfo {author} {\bibnamefont {et~al.}},\ }\bibfield  {title}
  {\bibinfo {title} {Observation of a quantum phase transition in the quantum
  rabi model with a single trapped ion},\ }\href@noop {} {\bibfield  {journal}
  {\bibinfo  {journal} {Nat. Commun.}\ }\textbf {\bibinfo {volume} {12}},\
  \bibinfo {pages} {1126} (\bibinfo {year} {2021})}\BibitemShut {NoStop}%
\bibitem [{\citenamefont {Hwang}\ and\ \citenamefont
  {Plenio}(2016)}]{PhysRevLett.117.123602}%
  \BibitemOpen
  \bibfield  {author} {\bibinfo {author} {\bibfnamefont {M.-J.}\ \bibnamefont
  {Hwang}}\ and\ \bibinfo {author} {\bibfnamefont {M.~B.}\ \bibnamefont
  {Plenio}},\ }\bibfield  {title} {\bibinfo {title} {Quantum phase transition
  in the finite jaynes-cummings lattice systems},\ }\href
  {https://doi.org/10.1103/PhysRevLett.117.123602} {\bibfield  {journal}
  {\bibinfo  {journal} {Phys. Rev. Lett.}\ }\textbf {\bibinfo {volume} {117}},\
  \bibinfo {pages} {123602} (\bibinfo {year} {2016})}\BibitemShut {NoStop}%
\bibitem [{\citenamefont {Jiang}\ \emph {et~al.}(2021)\citenamefont {Jiang},
  \citenamefont {Lu}, \citenamefont {Han}, \citenamefont {Fang}, \citenamefont
  {Zhao}, \citenamefont {Ma}, \citenamefont {Guo},\ and\ \citenamefont
  {Lee}}]{PhysRevA.104.043307}%
  \BibitemOpen
  \bibfield  {author} {\bibinfo {author} {\bibfnamefont {X.}~\bibnamefont
  {Jiang}}, \bibinfo {author} {\bibfnamefont {B.}~\bibnamefont {Lu}}, \bibinfo
  {author} {\bibfnamefont {C.}~\bibnamefont {Han}}, \bibinfo {author}
  {\bibfnamefont {R.}~\bibnamefont {Fang}}, \bibinfo {author} {\bibfnamefont
  {M.}~\bibnamefont {Zhao}}, \bibinfo {author} {\bibfnamefont {Z.}~\bibnamefont
  {Ma}}, \bibinfo {author} {\bibfnamefont {T.}~\bibnamefont {Guo}},\ and\
  \bibinfo {author} {\bibfnamefont {C.}~\bibnamefont {Lee}},\ }\bibfield
  {title} {\bibinfo {title} {Universal dynamics of the superradiant phase
  transition in the anisotropic quantum rabi model},\ }\href
  {https://doi.org/10.1103/PhysRevA.104.043307} {\bibfield  {journal} {\bibinfo
   {journal} {Phys. Rev. A}\ }\textbf {\bibinfo {volume} {104}},\ \bibinfo
  {pages} {043307} (\bibinfo {year} {2021})}\BibitemShut {NoStop}%
\bibitem [{\citenamefont {Chen}\ \emph
  {et~al.}(2020{\natexlab{b}})\citenamefont {Chen}, \citenamefont {Xie},\ and\
  \citenamefont {Chen}}]{PhysRevA.102.063721}%
  \BibitemOpen
  \bibfield  {author} {\bibinfo {author} {\bibfnamefont {X.-Y.}\ \bibnamefont
  {Chen}}, \bibinfo {author} {\bibfnamefont {Y.-F.}\ \bibnamefont {Xie}},\ and\
  \bibinfo {author} {\bibfnamefont {Q.-H.}\ \bibnamefont {Chen}},\ }\bibfield
  {title} {\bibinfo {title} {Quantum criticality of the rabi-stark model at
  finite frequency ratios},\ }\href
  {https://doi.org/10.1103/PhysRevA.102.063721} {\bibfield  {journal} {\bibinfo
   {journal} {Phys. Rev. A}\ }\textbf {\bibinfo {volume} {102}},\ \bibinfo
  {pages} {063721} (\bibinfo {year} {2020}{\natexlab{b}})}\BibitemShut
  {NoStop}%
\bibitem [{\citenamefont {Shen}\ \emph {et~al.}(2017)\citenamefont {Shen},
  \citenamefont {Yang}, \citenamefont {Wu},\ and\ \citenamefont
  {Zheng}}]{PhysRevA.95.013819}%
  \BibitemOpen
  \bibfield  {author} {\bibinfo {author} {\bibfnamefont {L.-T.}\ \bibnamefont
  {Shen}}, \bibinfo {author} {\bibfnamefont {Z.-B.}\ \bibnamefont {Yang}},
  \bibinfo {author} {\bibfnamefont {H.-Z.}\ \bibnamefont {Wu}},\ and\ \bibinfo
  {author} {\bibfnamefont {S.-B.}\ \bibnamefont {Zheng}},\ }\bibfield  {title}
  {\bibinfo {title} {Quantum phase transition and quench dynamics in the
  anisotropic rabi model},\ }\href {https://doi.org/10.1103/PhysRevA.95.013819}
  {\bibfield  {journal} {\bibinfo  {journal} {Phys. Rev. A}\ }\textbf {\bibinfo
  {volume} {95}},\ \bibinfo {pages} {013819} (\bibinfo {year}
  {2017})}\BibitemShut {NoStop}%
\bibitem [{\citenamefont {Zhang}\ \emph {et~al.}(2021)\citenamefont {Zhang},
  \citenamefont {Hu}, \citenamefont {Fu}, \citenamefont {Luo}, \citenamefont
  {Pu},\ and\ \citenamefont {Zhang}}]{PhysRevLett.127.063602}%
  \BibitemOpen
  \bibfield  {author} {\bibinfo {author} {\bibfnamefont {Y.-Y.}\ \bibnamefont
  {Zhang}}, \bibinfo {author} {\bibfnamefont {Z.-X.}\ \bibnamefont {Hu}},
  \bibinfo {author} {\bibfnamefont {L.}~\bibnamefont {Fu}}, \bibinfo {author}
  {\bibfnamefont {H.-G.}\ \bibnamefont {Luo}}, \bibinfo {author} {\bibfnamefont
  {H.}~\bibnamefont {Pu}},\ and\ \bibinfo {author} {\bibfnamefont {X.-F.}\
  \bibnamefont {Zhang}},\ }\bibfield  {title} {\bibinfo {title} {Quantum phases
  in a quantum rabi triangle},\ }\href
  {https://doi.org/10.1103/PhysRevLett.127.063602} {\bibfield  {journal}
  {\bibinfo  {journal} {Phys. Rev. Lett.}\ }\textbf {\bibinfo {volume} {127}},\
  \bibinfo {pages} {063602} (\bibinfo {year} {2021})}\BibitemShut {NoStop}%
\bibitem [{\citenamefont {Fallas~Padilla}\ \emph {et~al.}(2022)\citenamefont
  {Fallas~Padilla}, \citenamefont {Pu}, \citenamefont {Cheng},\ and\
  \citenamefont {Zhang}}]{PhysRevLett.129.183602}%
  \BibitemOpen
  \bibfield  {author} {\bibinfo {author} {\bibfnamefont {D.}~\bibnamefont
  {Fallas~Padilla}}, \bibinfo {author} {\bibfnamefont {H.}~\bibnamefont {Pu}},
  \bibinfo {author} {\bibfnamefont {G.-J.}\ \bibnamefont {Cheng}},\ and\
  \bibinfo {author} {\bibfnamefont {Y.-Y.}\ \bibnamefont {Zhang}},\ }\bibfield
  {title} {\bibinfo {title} {Understanding the quantum rabi ring using
  analogies to quantum magnetism},\ }\href
  {https://doi.org/10.1103/PhysRevLett.129.183602} {\bibfield  {journal}
  {\bibinfo  {journal} {Phys. Rev. Lett.}\ }\textbf {\bibinfo {volume} {129}},\
  \bibinfo {pages} {183602} (\bibinfo {year} {2022})}\BibitemShut {NoStop}%
\bibitem [{\citenamefont {Li}\ \emph {et~al.}()\citenamefont {Li},
  \citenamefont {Feng}, \citenamefont {Dai},\ and\ \citenamefont
  {Zhang}}]{zhang2023}%
  \BibitemOpen
  \bibfield  {author} {\bibinfo {author} {\bibfnamefont {L.-J.}\ \bibnamefont
  {Li}}, \bibinfo {author} {\bibfnamefont {L.-L.}\ \bibnamefont {Feng}},
  \bibinfo {author} {\bibfnamefont {J.-H.}\ \bibnamefont {Dai}},\ and\ \bibinfo
  {author} {\bibfnamefont {Y.-Y.}\ \bibnamefont {Zhang}},\ }\bibfield  {title}
  {\bibinfo {title} {Quantum rabi hexagonal ring in an artificial magnetic
  field},\ }\href@noop {} {\bibinfo  {journal} {arXiv:2304.01535}\
  }\BibitemShut {NoStop}%
\bibitem [{\citenamefont {Masson}\ \emph {et~al.}(2017)\citenamefont {Masson},
  \citenamefont {Barrett},\ and\ \citenamefont
  {Parkins}}]{PhysRevLett.119.213601}%
  \BibitemOpen
\bibfield  {journal} {  }\bibfield  {author} {\bibinfo {author} {\bibfnamefont
  {S.~J.}\ \bibnamefont {Masson}}, \bibinfo {author} {\bibfnamefont {M.~D.}\
  \bibnamefont {Barrett}},\ and\ \bibinfo {author} {\bibfnamefont
  {S.}~\bibnamefont {Parkins}},\ }\bibfield  {title} {\bibinfo {title} {Cavity
  qed engineering of spin dynamics and squeezing in a spinor gas},\ }\href
  {https://doi.org/10.1103/PhysRevLett.119.213601} {\bibfield  {journal}
  {\bibinfo  {journal} {Phys. Rev. Lett.}\ }\textbf {\bibinfo {volume} {119}},\
  \bibinfo {pages} {213601} (\bibinfo {year} {2017})}\BibitemShut {NoStop}%
\bibitem [{\citenamefont {Krauth}\ \emph {et~al.}(1992)\citenamefont {Krauth},
  \citenamefont {Caffarel},\ and\ \citenamefont {Bouchaud}}]{PhysRevB.45.3137}%
  \BibitemOpen
  \bibfield  {author} {\bibinfo {author} {\bibfnamefont {W.}~\bibnamefont
  {Krauth}}, \bibinfo {author} {\bibfnamefont {M.}~\bibnamefont {Caffarel}},\
  and\ \bibinfo {author} {\bibfnamefont {J.-P.}\ \bibnamefont {Bouchaud}},\
  }\bibfield  {title} {\bibinfo {title} {Gutzwiller wave function for a model
  of strongly interacting bosons},\ }\href
  {https://doi.org/10.1103/PhysRevB.45.3137} {\bibfield  {journal} {\bibinfo
  {journal} {Phys. Rev. B}\ }\textbf {\bibinfo {volume} {45}},\ \bibinfo
  {pages} {3137} (\bibinfo {year} {1992})}\BibitemShut {NoStop}%
\bibitem [{\citenamefont {Buonsante}\ \emph {et~al.}(2009)\citenamefont
  {Buonsante}, \citenamefont {Massel}, \citenamefont {Penna},\ and\
  \citenamefont {Vezzani}}]{PhysRevA.79.013623}%
  \BibitemOpen
  \bibfield  {author} {\bibinfo {author} {\bibfnamefont {P.}~\bibnamefont
  {Buonsante}}, \bibinfo {author} {\bibfnamefont {F.}~\bibnamefont {Massel}},
  \bibinfo {author} {\bibfnamefont {V.}~\bibnamefont {Penna}},\ and\ \bibinfo
  {author} {\bibfnamefont {A.}~\bibnamefont {Vezzani}},\ }\bibfield  {title}
  {\bibinfo {title} {Gutzwiller approach to the bose-hubbard model with random
  local impurities},\ }\href {https://doi.org/10.1103/PhysRevA.79.013623}
  {\bibfield  {journal} {\bibinfo  {journal} {Phys. Rev. A}\ }\textbf {\bibinfo
  {volume} {79}},\ \bibinfo {pages} {013623} (\bibinfo {year}
  {2009})}\BibitemShut {NoStop}%
\bibitem [{\citenamefont {Fisher}\ and\ \citenamefont
  {Barber}(1972)}]{PhysRevLett.28.1516}%
  \BibitemOpen
  \bibfield  {author} {\bibinfo {author} {\bibfnamefont {M.~E.}\ \bibnamefont
  {Fisher}}\ and\ \bibinfo {author} {\bibfnamefont {M.~N.}\ \bibnamefont
  {Barber}},\ }\bibfield  {title} {\bibinfo {title} {Scaling theory for
  finite-size effects in the critical region},\ }\href
  {https://doi.org/10.1103/PhysRevLett.28.1516} {\bibfield  {journal} {\bibinfo
   {journal} {Phys. Rev. Lett.}\ }\textbf {\bibinfo {volume} {28}},\ \bibinfo
  {pages} {1516} (\bibinfo {year} {1972})}\BibitemShut {NoStop}%
\bibitem [{\citenamefont {Botet}\ \emph {et~al.}(1982)\citenamefont {Botet},
  \citenamefont {Jullien},\ and\ \citenamefont {Pfeuty}}]{PhysRevLett.49.478}%
  \BibitemOpen
  \bibfield  {author} {\bibinfo {author} {\bibfnamefont {R.}~\bibnamefont
  {Botet}}, \bibinfo {author} {\bibfnamefont {R.}~\bibnamefont {Jullien}},\
  and\ \bibinfo {author} {\bibfnamefont {P.}~\bibnamefont {Pfeuty}},\
  }\bibfield  {title} {\bibinfo {title} {Size scaling for infinitely
  coordinated systems},\ }\href {https://doi.org/10.1103/PhysRevLett.49.478}
  {\bibfield  {journal} {\bibinfo  {journal} {Phys. Rev. Lett.}\ }\textbf
  {\bibinfo {volume} {49}},\ \bibinfo {pages} {478} (\bibinfo {year}
  {1982})}\BibitemShut {NoStop}%
\bibitem [{\citenamefont {Liu}\ \emph {et~al.}(2009)\citenamefont {Liu},
  \citenamefont {Zhang}, \citenamefont {Chen},\ and\ \citenamefont
  {Wang}}]{PhysRevA.80.023810}%
  \BibitemOpen
  \bibfield  {author} {\bibinfo {author} {\bibfnamefont {T.}~\bibnamefont
  {Liu}}, \bibinfo {author} {\bibfnamefont {Y.-Y.}\ \bibnamefont {Zhang}},
  \bibinfo {author} {\bibfnamefont {Q.-H.}\ \bibnamefont {Chen}},\ and\
  \bibinfo {author} {\bibfnamefont {K.-L.}\ \bibnamefont {Wang}},\ }\bibfield
  {title} {\bibinfo {title} {Large-$n$ scaling behavior of the ground-state
  energy, fidelity, and the order parameter in the dicke model},\ }\href
  {https://doi.org/10.1103/PhysRevA.80.023810} {\bibfield  {journal} {\bibinfo
  {journal} {Phys. Rev. A}\ }\textbf {\bibinfo {volume} {80}},\ \bibinfo
  {pages} {023810} (\bibinfo {year} {2009})}\BibitemShut {NoStop}%
\end{thebibliography}%

\end{document}